\documentclass[fleqn,usenatbib]{mnras}

\usepackage{newtxtext,newtxmath}
\usepackage[T1]{fontenc}
\usepackage[normalem]{ulem}

\DeclareRobustCommand{\VAN}[3]{#2}
\let\VANthebibliography\thebibliography
\def\thebibliography{\DeclareRobustCommand{\VAN}[3]{##3}\VANthebibliography}

\usepackage[dvipsnames]{xcolor}

\usepackage{graphicx}	
\usepackage{amsmath}	
\usepackage{pdflscape}
\usepackage{orcidlink}
\usepackage{soul}

\usepackage{acro}
\acsetup{
  make-links = false, 
  single = {1},
  single-style = long-short,
  list/display = all
}
\DeclareAcronym{PA}{short=PA,long=Position Angle}
\DeclareAcronym{DHOST}{short=DHOST,long=Degenerate Higher-Order Scalar-Tensor}
\DeclareAcronym{SOAR}{short=SOAR,long=SOuthern Astrophysical Research Telescope}
\DeclareAcronym{AO}{short=AO,long=Adaptative Optics}
\DeclareAcronym{AGNs}{short=AGNs,long=Active Galactic Nucleus}
\DeclareAcronym{eBOSS}{short=eBOSS,long=Extended Baryon Oscillation Spectroscopic Survey}
\DeclareAcronym{BOSS}{short=BOSS,long=Baryon Oscillation Spectroscopic Survey}
\DeclareAcronym{CNN}{short=CNN,long=Convolutional Neural Network}
\DeclareAcronym{CPU}{short=CPU,long=Central Processing Unit}
\DeclareAcronym{LPT}{short=LPT,long=Lagrangian Perturbation Theory}
\DeclareAcronym{CIC}{short=CIC,long=Cloud-in-a-Cell}
\DeclareAcronym{GPU}{short=GPU,long=Graphics Processing Unit}
\DeclareAcronym{IO}{short=I/O,long=Input/Output}
\DeclareAcronym{BAO}{short=BAO,long=Baryon Acoustic Oscillations}
\DeclareAcronym{2dFGRS}{short=2dFGRS,long=2dF Galaxy Redshift Survey}
\DeclareAcronym{2MRS}{short=2MRS,long=2MASS Redshift Survey}
\DeclareAcronym{2SLAQ}{short=2SLAQ,long=2dF-SDSS LRG and QSO Survey}
\DeclareAcronym{6dFGS}{short=6dFGS,long=6dF Galaxy Survey}
\DeclareAcronym{AGES}{short=AGES,long=AGN and Galaxy Evolution Survey}
\DeclareAcronym{BASS}{short=BASS,long=Beijing-Arizona Sky Survey}
\DeclareAcronym{CASLEO}{short=CASLEO,long=Complejo Astronómico El Leoncito}
\DeclareAcronym{CASTLES}{short=CASTLES,long=Cfa-Arizona-(H)ST-Lens-Survey}
\DeclareAcronym{CDM}{short=CDM,long=cold dark matter}
\DeclareAcronym{CfA}{short=CfA,long=Center for Astrophysics Redshift Survey}
\DeclareAcronym{CFHTLS}{short=CFHTLS,long=Canada-France-Hawaii Telescope Legacy Survey}
\DeclareAcronym{CFHT}{short=CFHT,long=Canada-France-Hawaii Telescope}
\DeclareAcronym{CNOC2}{short=CNOC2,long=Canadian Network for Observational Cosmology Field Galaxy Redshift Survey}
\DeclareAcronym{COLA}{short=COLA,long=COmoving Lagrangian Acceleration}
\DeclareAcronym{CS82}{short=CS82,long=CFTH/MegaCam Stripe-82 Survey}
\DeclareAcronym{DECaLS}{short=DECaLS,long=Dark Energy Camera Legacy Survey}
\DeclareAcronym{DECam}{short=DECam,long=Dark Energy Camera}
\DeclareAcronym{DEEP2}{short=DEEP2,long=DEEP2 Galaxy Redshift Survey}
\DeclareAcronym{DELVE}{short=DELVE,long=DECam Local Volume Exploration Survey}
\DeclareAcronym{DES}{short=DES,long=Dark Energy Survey}
\DeclareAcronym{ETG}{short=ETG,long=Early-Type Galaxies}
\DeclareAcronym{FLRW}{short=FLRW,long=Friedmann-Lemaître-Robertson-Walker}
\DeclareAcronym{FNN}{short=FNN,long=Fiducial Neural Network}
\DeclareAcronym{GAMA}{short=GAMA,long=Galaxy And Mass Assembly Survey}
\DeclareAcronym{GMOS}{short=GMOS,long=Gemini Multi-Object Spectrographs}
\DeclareAcronym{GNIRS}{short=GNIRS,long=Gemini Near-Infrared Spectrometer}
\DeclareAcronym{GR}{short=GR,long=General Relativity}
\DeclareAcronym{HSCSSP}{short=HSC-SSP,long=Hyper Suprime-Cam Subaru Strategic Program}
\DeclareAcronym{LCDM}{short=$\Lambda$CDM,long=$\Lambda$-Cold Dark Matter}
\DeclareAcronym{RGB}{short=RGB,long=Red--Green--Blue}
\DeclareAcronym{PSF}{short=PSF,long=Point Spread Function}
\DeclareAcronym{SN}{short=S/N,long=Signal to Noise}
\DeclareAcronym{RA}{short=RA,long=Right Ascension}
\DeclareAcronym{Dec}{short=Dec,long=Declination}
\DeclareAcronym{HSCLA}{short=HSCLA,long=Hyper Suprime-Cam Legacy Archive}
\DeclareAcronym{FITS}{short=FITS,long=Flexible Image Transport System}
\DeclareAcronym{SQL}{short=SQL,long=Structured Query Language}
\DeclareAcronym{HSC}{short=HSC,long=Hyper Suprime-Cam}
\DeclareAcronym{HST}{short=HST,long=Hubble Space Telescope}
\DeclareAcronym{IFU}{short=IFU,long=Integral Field Spectroscopic Units}
\DeclareAcronym{ISI}{short=ISI,long=Isolated Singular Isothermal}
\DeclareAcronym{JAIR}{short=JAIR,long=Joint Arc sample for Investigations into Relativity}
\DeclareAcronym{JWST}{short=JWST,long=James Webb Space Telescope}
\DeclareAcronym{KiDS}{short=KiDS,long=Kilo Degree Survey}
\DeclareAcronym{WiggleZ}{short=WiggleZ,long=WiggleZ Dark Energy Survey}
\DeclareAcronym{zCOSMOS}{short=zCOSMOS,long=zCOSMOS Program}
\DeclareAcronym{CFHTLenS}{short=CFHTLenS,long=Canada-France-Hawaii Telescope Lensing Survey}
\DeclareAcronym{RCSLenS}{short=RCSLenS,long=Red Cluster Sequence Lensing Survey}
\DeclareAcronym{RCS2}{short=RCS-2,long=Red-sequence Cluster Survey}
\DeclareAcronym{LAMOST}{short=LAMOST,long=Large Sky Area Multi-Object Fiber Spectroscopic Telescope Survey}
\DeclareAcronym{LaStBeRu}{short=LaStBeRu,long=Last Stand Before Rubin}
\DeclareAcronym{SIA}{short=SIA,long=Simple Image Access}
\DeclareAcronym{DECaPS}{short=DECaPS,long=DECam Plane Survey}
\DeclareAcronym{unWISE}{short=unWISE,long=unWISE Infrared Survey Explorer}
\DeclareAcronym{LCRS}{short=LCRS,long=Las Campanas Redshift Survey}
\DeclareAcronym{Legacy}{short=Legacy,long=DESI Legacy Imaging Surveys}
\DeclareAcronym{LinKS}{short=LinKS,long=Lenses in the Kilo-Degree Survey}
\DeclareAcronym{LOS}{short=LOS,long=line-of-sight}
\DeclareAcronym{SWEELS}{short=SWEELS,long=Sloan WFC Edge-on Late-type Lens Survey}
\DeclareAcronym{PDF}{short=PDF,long=Probability Distribution Function}
\DeclareAcronym{FWHM}{short=FWHM,long=Full Width at Half Maximum}
\DeclareAcronym{LSB}{short=LSB,long=DESI Legacy Sky Browser}
\DeclareAcronym{LSST}{short=LSST,long=Legacy Survey of Space and Time}
\DeclareAcronym{MCMC}{short=MCMC,long=Markov Chain Monte Carlo}
\DeclareAcronym{MGC}{short=MGC,long=Millennium Galaxy Catalogue}
\DeclareAcronym{MIRI}{short=MIRI,long=Mid-Infrared Instrument}
\DeclareAcronym{MLD}{short=MLD,long=Master Lens Database}
\DeclareAcronym{ML}{short=ML,long=Machine Learning}
\DeclareAcronym{MSE}{short=MSE,long=Mean Squared Error}
\DeclareAcronym{MUSE}{short=MUSE,long=Multi Unit Spectroscopic Explorer}
\DeclareAcronym{MzLS}{short=MzLS,long=Mayall z-band Legacy Survey}
\DeclareAcronym{NFW}{short=NFW,long=Navarro–Frenk–White}
\DeclareAcronym{NIFS}{short=NIFS,long=Near-Infrared Integral Field Spectrometer}
\DeclareAcronym{NN}{short=NN,long=Neural Network}
\DeclareAcronym{OzDES}{short=OzDES,long=Australian Dark Energy Survey}
\DeclareAcronym{QSO}{short=QSO,long=Quasi-Stellar Object}
\DeclareAcronym{Pan-STARRS}{short=Pan-STARRS,long=Panoramic Survey Telescope and Rapid Response System}
\DeclareAcronym{PPN}{short=PPN,long=Parameterized Post-Newtonian}
\DeclareAcronym{PSCz}{short=PSCz,long=PSCz Redshift Survey}
\DeclareAcronym{ReLU}{short=ReLU,long=Leaky Rectified Linear Unit}
\DeclareAcronym{REOSC}{short=REOSC,long=Recherche et Étude en Optique et Sciences Connexes}
\DeclareAcronym{SDSS}{short=SDSS,long=Sloan Digital Sky Survey}
\DeclareAcronym{SEP}{short=SEP,long=Source Extraction and Photometry}
\DeclareAcronym{SIEP}{short=SIEP,long=Singular Isothermal with Elliptic Potential}
\DeclareAcronym{SIE}{short=SIE,long=Singular Isothermal Ellipsoid}
\DeclareAcronym{SILO}{short=SILO,long=Spectroscopic Identification of Lensing Objects}
\DeclareAcronym{SIS}{short=SIS,long=Singular Isothermal Sphere}
\DeclareAcronym{SLACS}{short=SLACS,long=Sloan Lens ACS}
\DeclareAcronym{SL}{short=SL,long=Strong Lensing}
\DeclareAcronym{SNN}{short=SNN,long=Styled Neural Network}
\DeclareAcronym{SSRS}{short=SSRS,long=Southern Sky Redshift Survey}
\DeclareAcronym{SuGOHI}{short=SuGOHI,long=Survey of Gravitationally-lensed Objects in HSC Imaging}
\DeclareAcronym{VIPERS}{short=VIPERS,long=VIMOS Public Extragalactic Redshift Survey}
\DeclareAcronym{VLT}{short=VLT,long=Very Large Telescope}
\DeclareAcronym{VST}{short=VST,long=VLT Survey Telescope}
\DeclareAcronym{VVDS}{short=VVDS,long=VIMOS VLT Deep Survey}
\DeclareAcronym{ZA}{short=ZA,long=Zel'dovich approximation}
\DeclareAcronym{CCD}{short=CCD,long=Charge-Coupled Device}

\title[The Last Stand Before Rubin for Strong Lensing]{The Last Stand Before Rubin: a consolidated sample of strong lensing systems in wide-field surveys}

\author[Alves de Oliveira et al.]{
Renan Alves de Oliveira \orcidlink{0000-0002-0200-3833},$^{1}$\thanks{E-mail: \href{mailto:fisica.renan@gmail.com}{fisica.renan@gmail.com}}
João Paulo C. França \orcidlink{0009-0001-4409-6820},$^{2}$
Martín Makler \orcidlink{0000-0003-2206-2651},$^{3,2,1}$
\\
$^{1}$PPGCosmo, Centro de Ciências Exatas, Universidade Federal do Espírito Santo, Vitória, ES, 29075--910, Brazil\\
$^{2}$Centro Brasileiro de Pesquisas Físicas, Rio de Janeiro, RJ, 22290--180, Brazil\\
$^{3}$International Center for Advanced Studies \& Instituto de Ciencias Físicas, ECyT-UNSAM \& CONICET, San Martín, Buenos Aires, 1650, Argentina
}

\date{Accepted XXX. Received YYY; in original form ZZZ}

\pubyear{2025}

\begin{document}
\label{firstpage}
\pagerange{\pageref{firstpage}--\pageref{lastpage}}
\maketitle

\begin{abstract}
As the Vera Rubin Observatory begins its ten-year survey in 2025, it will probe key observables such as strong lensing (SL) by galaxies and clusters. In preparation for this new era, we assemble an extensive compilation of SL candidate systems from the literature, comprising over 30,000 unique objects that can be used as a watch list of known systems. By cross-matching this sample with photometric and spectroscopic catalogs, we construct two value-added tables containing key parameters for SL analysis, including lens and source redshifts and lens velocity dispersions $\sigma_v$. As a preparation for Rubin, we generate image cutouts for these systems in existing wide-field surveys with subarcsecond seeing, namely CFHTLens, CS82, RCSLens, KiDS, HSC, DES, and DESI Legacy. This sample, dubbed the ``Last Stand Before Rubin'' (LaStBeRu), has a myriad of applications, from using archival data to selections for follow-up projects and training of machine learning algorithms. As an application, we perform a test of General Relativity using these data, combining the effects of motion of massless particles (through SL modeling) and non-relativistic bodies through $\sigma_v$, which allow one to set constraints on the Post-Newtonian parameter $\gamma_\mathrm{PPN}$. Using the LaStBeRu database, we present an independent test of $\gamma_\mathrm{PPN}$ (distinct from previous analyses) and, for the first time, we present such a test exclusively with systems identifiable in ground-based images. By combining these data with the previously published samples, we obtain the most stringent constraint on $\gamma_\mathrm{PPN}$. Our results are consistent with GR at the $\sim$~1-$\sigma$ level and with the previous results from the literature.
\end{abstract}




\section{Introduction} \label{sec:intro}

\ac{SL} is a phenomenon predicted by \ac{GR} in which a massive object bends the light of a distant source when there is an alignment between them and the observer~\citep{SEF}. This effect has proven to be a powerful astrophysical tool with wide-ranging implications, such as characterizing high-redshift sources~\citep{anupreeta_2016, Borsato_24, Amvrosiadis_24}, constraining cosmological parameters~\citep{gavazzi_2008, Schwab_et_al_2010, 2017MNRAS.468.2590S, 2020MNRAS.498.1420W, millon_2020, shajib_2023}, probing dark matter properties~\citep{2007ApJ...667..176G, gleen_2009, montel_2022, wagner-carena_2023}, or characterizing the density profile in the lens galaxy~\citep{Auger_et_al._2009, bolton_2012, Sonnenfeld_et_al._2013_2, shajib_2019, nightingale_2019, gilman_2020, etherington_2023}. \ac{SL} systems have been observed in many astronomical surveys and have been extensively studied with follow-up observations~\citep{2023NatAs.tmp....7V}, as they are an ideal tool for understanding and testing gravity~\citep{Schwab_et_al_2010, Cao_et_al_2016, Liu_et_al_2022}.

However, finding \ac{SL} systems is challenging. This is particularly true for galaxy--galaxy strong lenses, the primary focus of this work, which have scales of arcseconds and are thus strongly affected by seeing in ground-based observations~\citep{2019A&A...625A.119M}. Discoveries rely on diverse methods, including morphology-based searches for features like arcs and rings~\citep{Allan_et_al._2006, 2018MNRAS.479..262G}, spectroscopic identification of high-redshift emission lines overprinting a foreground galaxy's spectrum, and the detection of high-luminosity submillimeter galaxies, among others~\citep{2006ApJ...638..703B, 2012ApJ...744...41B, Stark_et_al._2013, Talbot_et_al._2020}. Most of these methods require complementary data for confirmation and astrophysical application (e.g., spectra for image-based candidates, and vice-versa). For a comprehensive overview of search techniques, we refer the reader to the review by~\citet{2024SSRv..220...23L}.

\begin{figure*}
	\centering
	\includegraphics[width = \textwidth]{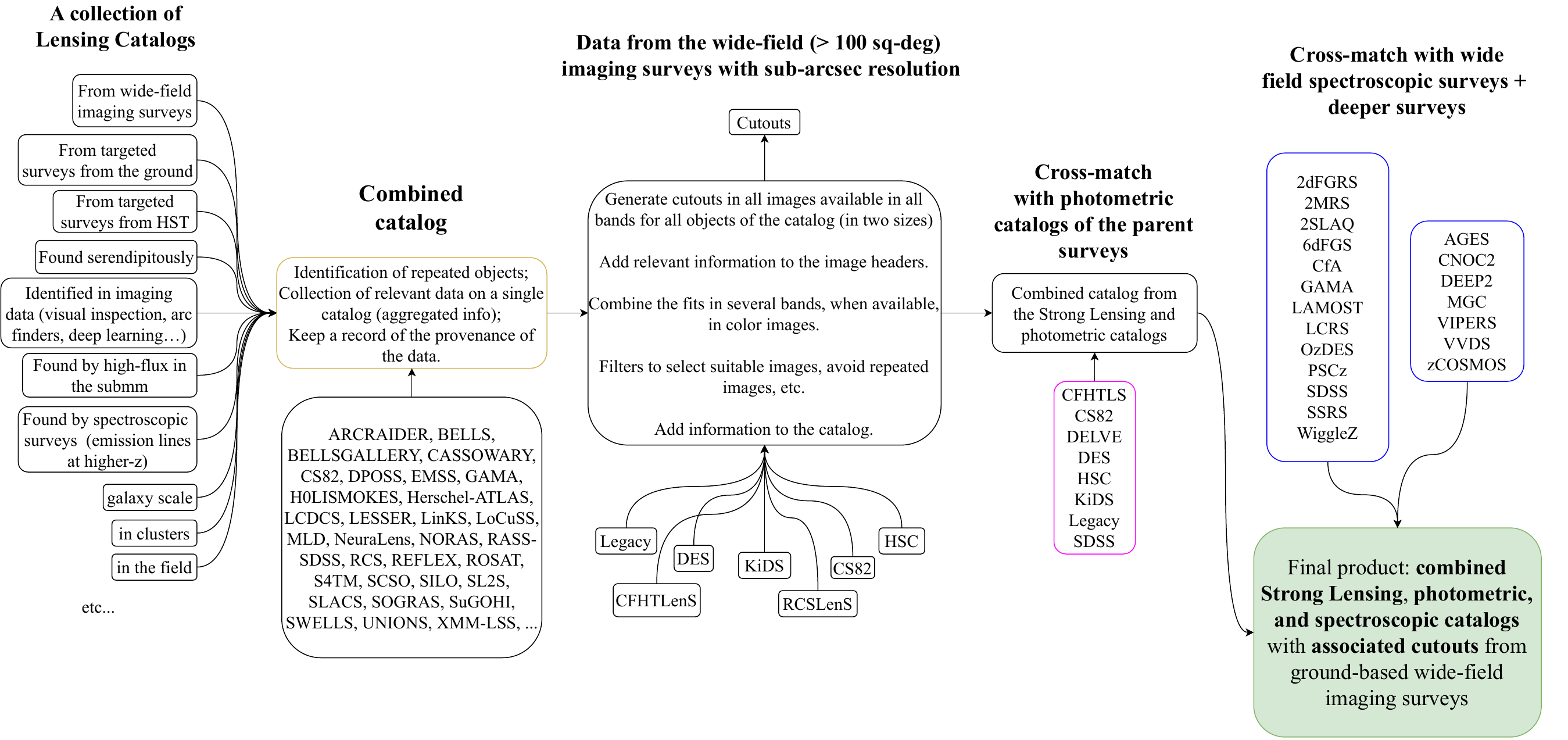}
	\caption{This image summarizes the steps to build the ``slcomp'' framework,  which aggregates many lensed systems into a single database.}
	\label{fig:framework}
\end{figure*}

While space-based missions like Euclid are discovering a large number of lenses~\citep{2025arXiv250209802N, euclid_colab_sl}, the Vera C. Rubin Observatory \ac{LSST} will provide the time-domain component and ground-based imaging data~\citep{LSST:2008ijt}. Besides, due to the angular scales, the Rubin lenses will be more suited for follow-up observations, which are mostly carried out from the ground (and subject to the same \ac{PSF} conditions as Rubin). On the ground-based side, we expect to find an increasing number of \ac{SL} systems, likely reaching $\mathcal{O}(10^5)$~\citep{collett_2015, 2019arXiv190205141V}, thanks to the high-quality, subarcsecond-seeing data from the Rubin Observatory. To improve the detection of these systems, precursor automated arcfinding techniques in imaging surveys have also led to the discovery of some \ac{SL} systems~\citep{2007ApJ...660.1176E, 2008AJ....135..664H, Diehl_et_al._2009, Diehl_et_al._2017, ODonnell_et_al._2021}. However, recent advances in \ac{ML} techniques in astronomy have led to robust simulations that improve arcfinders and identify several new \ac{SL} candidates~\citep{LinKS, Huang_et_al._2020,Canameras_et_al._2020}. Therefore, this also opens the door to reducing observational biases and the rate of false positives~\citep{2023arXiv231107455H}. The deep, multi-band, and high-cadence nature of the \ac{LSST} will also enable the creation of watchlists to systematically search for lensed transients (e.g., the Fink Broker~\citep{2021MNRAS.501.3272M}), such as supernovae and gamma-ray bursts.

Although \ac{SL} systems are rare objects, there are still plenty of lensing catalogs for wide-field surveys available, such as the \ac{MLD}\footnote{\url{http://admin.masterlens.org}}~\citep{Moustakas_et_al._2012}, or lenscat\footnote{\url{https://github.com/lenscat/lenscat}}~\citep{2024arXiv240604398V}. All these databases contain data from multiple systems, suitable for monitoring and improving observational features. However, images of the objects are not always available from that specific database; sometimes, no data are available from observations on the ground (e.g., only from space surveys). A unified resource that aggregates candidate systems from these different detection methods and systematically provides homogeneous imaging from ground-based precursor data to Rubin is still needed. Such a resource is essential to prepare for the \ac{LSST} data stream.

This work introduces the slcomp framework (Figure~\ref{fig:framework}) to address this gap. Our primary focus is on galaxy--galaxy \ac{SL} systems, although we also include galaxy clusters and lensed \ac{QSO}s on a best-effort basis. The main objective is to group together candidates from various sources, allowing for uniform visual inspection and cross-matching of photometric and spectroscopic data. We aim to build a comprehensive sample suitable for multiple applications, including: (i) creating a sort of ``SL ImageNet'' to train and fine-tune \ac{ML} algorithms; (ii) providing a consolidated sample for cosmological studies; (iii) aiding in the selection of targets for follow-up observations; and (iv) establishing watchlists for transient events.

In Section~\ref{sec:literature}, we describe how the lensed systems were compiled from the literature and the data were homogenized. In Section~\ref{sec:matches}, we present the matches between objects from the literature and spectroscopic surveys. We consolidate all tabular data in Section~\ref{sec:consolidated}. We then present a catalog that combines images from many wide-field surveys in Section~\ref{sec:cutouts}. In Section~\ref{sec:lastberu}, we present the slcomp framework and the resulting database, showcasing some applications in Section~\ref{sec:lastberu-applications}. Finally, Section~\ref{sec:conclusions} presents our concluding remarks.

\section{Consolidated SL Database} \label{sec:literature}
 
\subsection{Sources of the SL Catalogs} \label{subsec:slcatalogs}

The rise in ground-based observation surveys over the years, particularly those employing spectroscopic follow-up, led to a variety of catalogs of confirmed and candidate strong-lensing systems, such as the \ac{SLACS}~\citep{2006ApJ...638..703B}, the \ac{SuGOHI}~\citep{Sonnenfeld_et_al._2018}, the \ac{LinKS}~\citep{LinKS}, or the \ac{SILO}~\citep{Talbot_et_al._2020}. These catalogs provided data for confirmed and potential lensing systems, their respective images, and measured/modeled lensing features. However, \ac{SL} systems are also discovered through a variety of other methods, including imaging-based searches, submillimeter observations, and selections from clusters and \ac{QSO} catalogs, using data from both ground and space based telescopes.

The problem of combining data from many surveys and catalogs is nontrivial because each provides a unique, nonhomogenous data structure. For this work, we first compiled data from numerous collections of strong-lensed systems discovered through different methods, including serendipitous discoveries~\citep{Allan_et_al._2006, Bettinelli_et_al._2016, Tanaka_et_al._2016}, spectroscopic observations~\citep{Auger_et_al._2009, Bolton_et_al._2008, 2012ApJ...744...41B, Cao_et_al._2020, Feron_et_al._2009, Focardi_et_al._2015, Goobar_et_al._2017, Holwerda_et_al._2015, Jaelani_et_al._2021, Kubo_et_al._2008, 2010ApJ...724L.137K, Laseter_et_al._2022, Schwope_et_al._2010, Shu_et_al._2016, Shu_et_al._2017, Sonnenfeld_et_al._2013, Stark_et_al._2013, Talbot_et_al._2018, Talbot_et_al._2020, Talbot_et_al._2022, Treu_et_al._2011, Warren_et_al._1998}, submillimeter data~\citep{Amvrosiadis_et_al._2018, Negrello_et_al._2016}, X-ray clusters~\citep{Bohringer_et_al._2000, Bohringer_et_al._2004, Cypriano_et_al._2004, Ebeling_et_al._1998, Ebeling_et_al._2000, Gioia_et_al._1990, 2010A&A...513A...8K, Pierre_et_al._2006, Popesso_et_al._2004}, visual and automated inspection~\citep{2007A&A...461..813C, Chan_et_al._2020, CS82, Diehl_et_al._2009, Diehl_et_al._2017, 2008ApJS..176...19F, 2012arXiv1210.4136F, Geach_et_al._2015, Gladders_et_al._2005, Gonzalez_et_al._2001, 2008AJ....135..664H, Jaelani_et_al._2020, Lin_et_al._2008, Lopes_et_al._2005, Menanteau_et_al._2010, 2012ApJ...749...38M, Nord_et_al._2019, ODonnell_et_al._2021, Olsen_et_al._2006, Pawase_et_al._2014, 2005ApJ...627...32S, Sonnenfeld_et_al._2018, Sonnenfeld_et_al._2019, Sonnenfeld_et_al._2020, Sygnet_et_al._2010, van_Breukelen_et_al._2006, 2011RAA....11.1185W, Wong_et_al._2018, Wong_et_al._2022}, and using \ac{ML}~\citep{Canameras_et_al._2020, Canameras_et_al._2021, Huang_et_al._2020, Huang_et_al._2021, Jacobs_et_al._2018, Jacobs_et_al._2019, Li_et_al._2021, LinKS, Rojas_et_al._2022, Savary_et_al._2021, Shu_et_al._2022, Stein_et_al._2021, Storfer_et_al._2022, Zaborowski_et_al._2022}. In addition, we incorporated other compilations of \ac{SL} systems, such as~\citet{Cao_et_al._2015, Leaf_et_al._2018, Richard_et_al._2010, Sonnenfeld_et_al._2013_2, Chen_et_al_2019} and \ac{MLD}.

The \acl{MLD}~\citep{Moustakas_et_al._2012} has been, until recently, the most comprehensive collection of known lensed systems. This compilation features galaxy--galaxy, galaxy--quasar, and cluster--galaxy lenses, and includes various lensing parameters, such as lens and source redshifts, magnitudes, and Einstein radius, when available. For some systems, \ac{MLD} also provides lens models, particularly for those with high-resolution images from space-based telescopes like the \ac{HST}. Even though \ac{MLD} compiles data for a large number of lensed systems, there is no unified database containing multi-survey images and systematic photometric/spectroscopic cross-matches through surveys on a single interface.

\subsection{Collection and Combination of SL Data} \label{subsec:slfeatures}

Combining data from different catalogs presents the initial challenge of achieving uniformity. To address this, we must first establish a method to uniquely identify each \ac{SL} system across various sources. This allows us to match and consolidate all available information for what is, in fact, the same object on the sky. Our primary step is therefore to define a unique identifier based on the system's celestial coordinates.

Our unique identifier is the system's JNAME, which is based on its J2000.0 coordinates. The JNAME for each system is generated from the averaged \ac{RA} and \ac{Dec} coordinates derived from all available literature entries for that object. This averaged position then becomes the definitive positional reference for the entry in our catalog, formatted as JHHMMSS.s$\pm$DDMMSS.s (H: hour; M: minutes; S: seconds, with double precision; D: degrees), with tenth-of-an-arcsecond precision. This approach mitigates small discrepancies in coordinates from different sources, which could otherwise lead to inconsistencies when cross-matching catalogs.

To group entries corresponding to the same system, we perform a positional cross-match. We consider objects to be part of the same system if their coordinates fall within a specific search radius. To determine an appropriate radius, we first performed an internal cross-match within each catalog listed in Section~\ref{subsec:slcatalogs} using a radius of $10^{\prime\prime}$. The size of this radius was motivated because it is half of the smallest cutout size from Section~\ref{sec:cutouts}. After performing this initial cross-match, we found some internal matches of objects of the same catalog in the literature: two matches in~\citet{Gonzalez_et_al._2001, Olsen_et_al._2006, Shu_et_al._2022}, four matches in~\citet{Holwerda_et_al._2015}, six matches in~\citet{Stein_et_al._2021}, eight matches in~\citet{Stark_et_al._2013}, 52 matches from the \ac{SuGOHI} sample in~\citet{Chan_et_al._2020,Jaelani_et_al._2020,Jaelani_et_al._2021,Sonnenfeld_et_al._2018,Sonnenfeld_et_al._2019,Sonnenfeld_et_al._2020,Wong_et_al._2018}, 68 matches in~\citet{LinKS}, 110 matches in~\citet{Storfer_et_al._2022}, and 2852 matches in~\citet{Moustakas_et_al._2012}.

\begin{figure}
    \begin{center}
        \includegraphics[height=0.5\columnwidth]{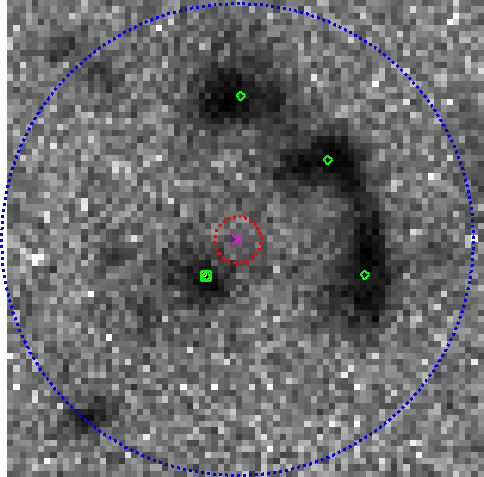}    
    \end{center}
    \caption{Single band image of the gravitational lens system J004106.3-415534.5 obtained by the Dark Energy Survey, reported in~\citet{Moustakas_et_al._2012, Diehl_et_al._2017, Nord_et_al._2019}, where the magenta cross at the center is the average of each system coordinate reported from the literature, representing the adopted JNAME coordinates.}\label{fig:matchingerror}
\end{figure}

The high number of internal matches within the \ac{MLD} data is expected, as \ac{MLD} itself is a meta-catalog that compiles entries from various other sources. This inherently leads to some objects being listed multiple times, occasionally with coordinates centered on different images of the lensed source rather than the lens itself. Our automated approach of using the average coordinate helps to unify these entries, though it can occasionally result in a central position that does not perfectly align with the lens galaxy's center. These numbers also reveal the presence of clusters, pairs of interacting galaxies (e.g., mixed objects presented in~\citet{Holwerda_et_al._2015}), and a combination of clusters and galaxies. Therefore, we assign the following rules to generate a unique JNAME:

\begin{enumerate}
    \item Objects with a separation of less than one arc-second receive the same JNAME.
    \item Objects matched within a radius between 1 and 10 arc-seconds also receive the same JNAME.~\label{radius10}
    \item Objects with no match within a ten-arc-second radius are designated as single systems.
\end{enumerate}

With a unique JNAME assigned to each system, we proceeded to consolidate the available data. To build a homogeneous dataset, we selected a set of key physical parameters that we consider relevant for lensing studies and were commonly available, when possible, across the different sources. These parameters include: the original catalog identification (Reference), lens and source redshifts, velocity dispersion, Einstein radius, classification rank, system types, and magnitudes in the \ac{SDSS} bands $u$, $g$, $r$, $i$, $z$, and $y$. Also, since many systems have magnitudes from the \ac{HST} filter $F814W$, we also keep this band.

In Section~\ref{sec:matches}, we matched all subjects in the database with photometric and spectroscopic catalogs using the coordinates extracted from JNAME. Figure~\ref{fig:matchingerror} shows the impact of assigning a unique JNAME to objects originating from the item~\ref{radius10}, where each green diamond represents the individual coordinate of their respective catalog, the magenta cross is the positional average of these green diamonds, derived from the positional coordinates of JNAME. Red and blue circles with radii of $1^{\prime\prime}$ and $10^{\prime\prime}$, respectively, are shown. However, as this decision only impacts the cross-matching of objects, and our main objective in this catalog is to study single-galaxy lenses, it does not necessarily present a dataset problem.

Some references report information related to each type of system, for example, systems composed of cluster--galaxy, galaxy--galaxy, cluster--quasar, and group--galaxy, where the first object of each pair indicates the lens type and the second slot is the source type. In addition, some references provide columns for the lens and source types. Therefore, we added two features to the dataset that represent these classifications, namely ``Lens Type'' and ``Source Type''. We assigned the ``System Type'' according to two criteria: (i) the lens and source types reported in the literature, and (ii) our JNAME assignment rules. If the ``Lens Type'' is a galaxy or a quasar, the ``System Type'' is ``Single Lens Galaxy''. For clusters, X-ray clusters, or groups of galaxies, the ``System Type'' is ``Non Single Lens Galaxy''. These classifications are convenient for distinguishing single-galaxy lenses from more complex configurations, such as systems with two nearby lensing galaxies or group/cluster-scale lenses. For systems with the same JNAME that were merged under rule~\ref{radius10}, we classified the ``System Type'' as ``Non-Single Lens Galaxy Merged'', emphasizing that this entry might represent a component of a cluster or a binary system where coordinates of multiple components were merged into a single identifier. If there were inconsistencies in the classification of a system across different references, we leave this field empty.

Spectroscopic measurements of redshifts are generally more accurate and precise than photometric observations. Therefore, we prioritize spectroscopic measurements for redshift values. When a reference provides both spectroscopic ($z_{L,S}$Type=S) and photometric ($z_{L,S}$Type=P) redshifts, we retain only the spectroscopic one, along with its error ($\delta z_{L,S}$) and source reference ($z_{L,S}$Ref). If no redshift nature type is specified, we keep the reference value\footnote{NaN means ``Not a Number'', and it is a null value.} ($z_{L,S}$Type=NaN). Additional information related to spectroscopy includes, when available in the source catalog, the lens velocity dispersion ($\sigma_v$) and its corresponding observational error ($\delta \sigma_v$), along with the source reference ($\sigma_v$Ref).

The Einstein radius ($\theta_E$) is another essential feature we collect, as it contains information related to the total mass of the lens. However, since it is a highly model-dependent feature, we also record the method used to derive it in the column $\theta_E$Method. This allows one to differentiate, for example, values obtained from detailed gravitational lens modeling (e.g., using power-law or composite mass models) from those derived from rough estimators, such as the curvature radius of arcs or the separation between multiple images. Acknowledging the complexity, we store the reported derivation method to provide context for the $\theta_E$ value.

A significant challenge arises from the variety of ways magnitudes are reported. Magnitudes can differ based on the photometric system (e.g., AB, Vega), the filter used (e.g., \ac{SDSS} bands, Johnson-Cousins bands), and the measurement definition (e.g., Kron, fixed aperture, etc.). To avoid an unmanageable number of columns, we adopted a naming convention to map the vast set of possible definitions into a few key columns, focusing on optical bands relevant for current and future surveys like the \ac{LSST}. We associate magnitudes from different filter systems to the closest corresponding \ac{SDSS} band, following, for example, Figure 1 of~\citet{bessel}:

\begin{itemize}
	\item $\textrm{mag}_V\rightarrow\textrm{mag}_r$
	\item $\textrm{mag}_B\rightarrow\textrm{mag}_g$
	\item $\textrm{mag}_R\rightarrow\textrm{mag}_i$
	\item $\textrm{mag}_I\rightarrow\textrm{mag}_z$
\end{itemize}

Other reported magnitudes were renamed as follows:

\begin{itemize}
	\item $\textrm{mag}_\textrm{rauto}\rightarrow\textrm{mag}_r$
	\item $\textrm{Megacam/i}\rightarrow\textrm{mag}_i$
	\item $\textrm{Megacam/g}\rightarrow\textrm{mag}_g$
	\item $\textrm{i}\rightarrow\textrm{mag}_i$
	\item $\textrm{r}\rightarrow\textrm{mag}_r$
	\item $\textrm{i+}\rightarrow\textrm{mag}_i$
\end{itemize}

Crucially, to preserve the original information, we store the provenance for each individual measurement in our consolidated catalog. As mentioned earlier, we did not include other Hubble magnitudes from \ac{HST} filters $F105W$, $F110W$, $F125W$, $F140W$, $F160W$, $F475X$ (WFC3), $F606W$ (ACS), $F606W$ (WFPC2), and the infrared bands $W$, $H$, $K$ due to the small number of systems with these bands (less than 30, in total, containing at least one of these bands), and given our focus on the optical, especially on the bands closer to the \ac{LSST} ones. In some cases, the uncertainties in the magnitudes are reported as asymmetric error bars. In these cases, we report a mean error bar, due to the database structure, as there is a single column for the magnitude errors.

Finally, we include in the database a grading for the quality of the \ac{SL} system from the original catalogs. This grade reflects the different mechanisms for classifying the candidates in these catalogs, from visual inspections to spectroscopic quality, and score from machine learning classifiers.

In this Section, we have compiled relevant information from various sources for objects that share the same JNAME. This extensive compilation is presented as a database, which can be used as a watchlist of confirmed and candidate \ac{SL} systems, constituting the first tabular data product of this work. The structure of our database is such that each unique system, identified by its JNAME, serves as a primary key. The corresponding record aggregates all collected data for that system from the various literature sources. Specifically, for each physical quantity (like redshift or a specific magnitude), the entry contains an array of all available values found in the literature, along with another array storing the corresponding reference for each value. This ensures that all original information and its provenance are preserved. Table~\ref{tab:dataset-description} describes all the columns used in this dataset.

\begin{table*}
    \centering
	\caption{Description of the columns used in the first table of the database -- the combined compilation of SL candidate systems. The first column shows the content of each variable in the catalog, while the second column shows the actual column names in the database. The third column presents shorts descriptions for the variables used in the dataset.}
	\label{tab:dataset-description}
	\resizebox{\textwidth}{!}{
    \begin{tabular}{lll}
			\hline
			Feature                                        & Dataset Key                   & Description                                                                      \\
			\hline
			Reference                                      & Reference                     & Source catalog/paper identifier.                                                 \\
			Unique Identifier                              & JNAME                         & Catalog unique identifier in HHMMSS.s+/-DDMMSS.s                                  \\
			Original Identifier                            & Original\_ID                  & Original identifier from the source catalog/paper.                               \\
			Alternative Name                               & Alternative\_Name             & Other name used in the source catalog/paper.                                     \\
			RA                                           & RA                            & JNAME right ascension coordinate (J2000).                                        \\
			Dec                                          & DEC                           & JNAME declination coordinates (J2000).                                           \\
			Individual Coordinates                         & Individual\_Coordinates       & Object coordinate as is from the source catalog/paper in HHMMSS.ss+\/-DDMMSS.ss. \\
			System Type                                    & System\_Type                  & Classification of the lensing system.                                            \\
			Lens/Source Type                               & \{Lens,Source\}\_Type         & Lens/Source type.                                                                \\
			$z_{L,S}$                                      & z\_\{L,S\}                    & Lens/Source redshift.                                                            \\
			$\delta z_{L,S}$                               & z\_\{L,S\}Err                 & Lens/Source redshift error.                                                      \\
			$z_{L,S}$ Type                                 & z\_\{L,S\}Type                & Lens/Source redshift type.                                                       \\
			$z_{L,S}$ Reference                            & z\_\{L,S\}Ref                 & Lens/Source redshift provenance.                                                 \\
			$\sigma_v$                                     & velDisp                       & Velocity dispersion (km/s).                                                      \\
			$\delta\sigma_v$                               & velDispErr                    & Velocity dispersion error (km/s).                                                \\
			$\sigma_v$ Reference                           & velDispRef                    & Velocity dispersion provenance.                                                  \\
			$\theta_E$                                     & theta\_E                      & Einstein radius (arcsec).                                                        \\
			$\delta\theta_E$                               & theta\_EErr                   & Einstein radius error (arcsec).                                                  \\
			$\theta_E$ Method                              & theta\_EMethod                & Method used to obtain the Einstein radius.                                       \\
			${\textrm{mag}}_{u,g,r,i,z,y,F814W}$           & mag\_\{u,g,r,i,z,y,F814W\}    & Magnitude in \{u,g,r,i,z,y,F814W\} band.                                         \\
			$\delta{\textrm{mag}}_{u,g,r,i,z,y,F814W}$     & mag\_\{u,g,r,i,z,y,F814W\}Err & Magnitude error in \{u,g,r,i,z,y,F814W\} band.                                   \\
			${\textrm{mag}}_{u,g,r,i,z,y,F814W}$ Reference & mag\_\{u,g,r,i,z,y,F814W\}Ref & Magnitude provenance for \{u,g,r,i,z,y,F814W\} band.                             \\
			${\textrm{mag}}_{u,g,r,i,z,F814W}^S$           & mag\_\{u,g,r,i,z,F814W\}S     & Source magnitude in \{u,g,r,i,z,F814W\} band.                                    \\
			${\textrm{mag}}_{u,g,r,i,z,F814W}^S$ Reference & mag\_\{u,g,r,i,z,F814W\}SRef  & Source magnitude provenance for \{u,g,r,i,z,F814W\} band.                        \\
			Other Matches                                  & Other\_Matches                & Object found in different catalogs, not necessarily in this dataset.        \\
			Original Comments                              & Original\_Comments            & Notes related to lensing features.     \\
			Grade                                          & Grade                         & Gradings from multiple methods. \\
			\hline
	\end{tabular}}
\end{table*}

\section{Cross-Matching with Photometric and Spectroscopic Catalogs} \label{sec:matches}

The previously introduced unique object identifier, JNAME, is appropriate for searching objects in other databases because of its coordinate representation. The main goal of this work is to expand the dataset for systems found in the literature, matching them with current data and combining them with other survey catalogs.

This choice for matching radius is motivated by the need to balance astrometric uncertainty with source density. A radius of $1^{\prime\prime}$ is large enough to encompass the typical astrometric uncertainty, thus avoiding the loss of true matches, but small enough to prevent false matches due to high source density. To ensure that each object had a unique counterpart in both catalogs, we employed the nearest-neighbor approach. While the small radius size minimizes the occurrence of multiple matches, this approach selects the closest object from the cross-matched catalog in the rare cases where multiple matches occur within the search radius. This method assumes that the lens is a single, centrally located galaxy, and thus the counterpart is the one closest to its center.

\subsection{Photometric Catalogs} \label{subsec:photomatches}

The motivation for using photometric data is the information available, such as redshifts, magnitudes, and morphological properties, that are easily available from many wide-field surveys (i.e., large footprints). Part of this information, which is not typically reported in \ac{SL} catalogs, is crucial for a wide range of applications (see Sections~\ref{sec:lastberu-applications-follow} and~\ref{sec:lastberu-applications-mod-grav}).

We obtained photometric tables from many wide-field surveys presented in Section~\ref{sec:cutouts}, including the final data release of the \ac{CFHTLS}~\citep{CFHTLS}, the \ac{CS82}~\citep{CS82}, the second data release of \ac{DELVE}~\citep{DELVE}, the second data release of \ac{DES}~\citep{DESDR2}, the third release of \ac{HSCSSP} (we refer to as \acs{HSC};~\citet{HSCDR3}), the third release of \ac{KiDS}~\citep{KiDSDR3}, the ninth data release of the DESI \ac{Legacy} Survey~\citep{Legacy} and the seventeenth release of \ac{SDSS}~\citep{SDSSDR17}.

To obtain photometric data from the \ac{CFHTLS} (Deep and Wide Fields) and \ac{KiDS} surveys, we used the CDS xMatch service\footnote{\url{http://cdsxmatch.u-strasbg.fr}}~\citep{2012ASPC..461..291B}, where we input the table built in Section~\ref{subsec:slfeatures}, and tables ``CFHTLS Wide'', ``CFHTLS Deep'', and ``KiDS DR3'' from VizieR~\citep{2000A&AS..143...23O}, using the radius of $1^{\prime\prime}$ and all-sky cross-match area. 

For the \ac{DELVE}, \ac{DES}, DESI \ac{Legacy}, and \ac{SDSS} surveys, we used the X-Match service provided by Astro Data Lab\footnote{\url{https://datalab.noirlab.edu}}. For \ac{DELVE}, we used the table ``objects''  and the value-added catalog ``photoz'' for extraction of magnitudes and photometric redshifts, respectively. We only used the table ``mag'' for the \ac{DES} survey. We used table ``tractor'' to obtain the magnitudes for all the systems in the survey footprint and to link the data with the value-added catalog ``photo\_z'' from the DESI \ac{Legacy} survey. Finally, for the \ac{SDSS} survey, we used only the ``phototable'' to obtain photometric magnitudes and morphological information. For all these cross-matches, we adopted a matching radius of $1^{\prime\prime}$ and the ``Nearest neighbor'' option.

Extracting data from the \ac{HSC} survey involved several steps. Initially, we generated a \ac{FITS} catalog that contained positional information for the dataset presented in Section~\ref{subsec:slfeatures}. This catalog was then used as input for the hscSspCrossMatch tool\footnote{\url{https://hsc-gitlab.mtk.nao.ac.jp/ssp-software}} to generate an \ac{SQL} query. Subsequently, we submitted this query to CAS Search\footnote{\url{https://hsc-release.mtk.nao.ac.jp/datasearch}}. Within the \ac{SQL} query, we match the dataset from the literature with the ``forced'' tables from the Wide and Deep/Ultra Deep fields, using the specified matching radius from earlier stages. 

With this match, we were able to link objects from the \ac{HSC} internal catalog, enabling the use of other value-added catalogs for redshifts from tables``photoz\_demp'', ``photoz\_dnnz'', and ``photoz\_mizuki''. We then filtered the redshift data by selecting, for each object, the estimate from the three methods with the lowest ``risk'' flag value (which indicates the probability of being incorrect or an outlier), thus providing a more robust photometric redshift estimation.

\begin{figure}
	\includegraphics[width=\columnwidth]{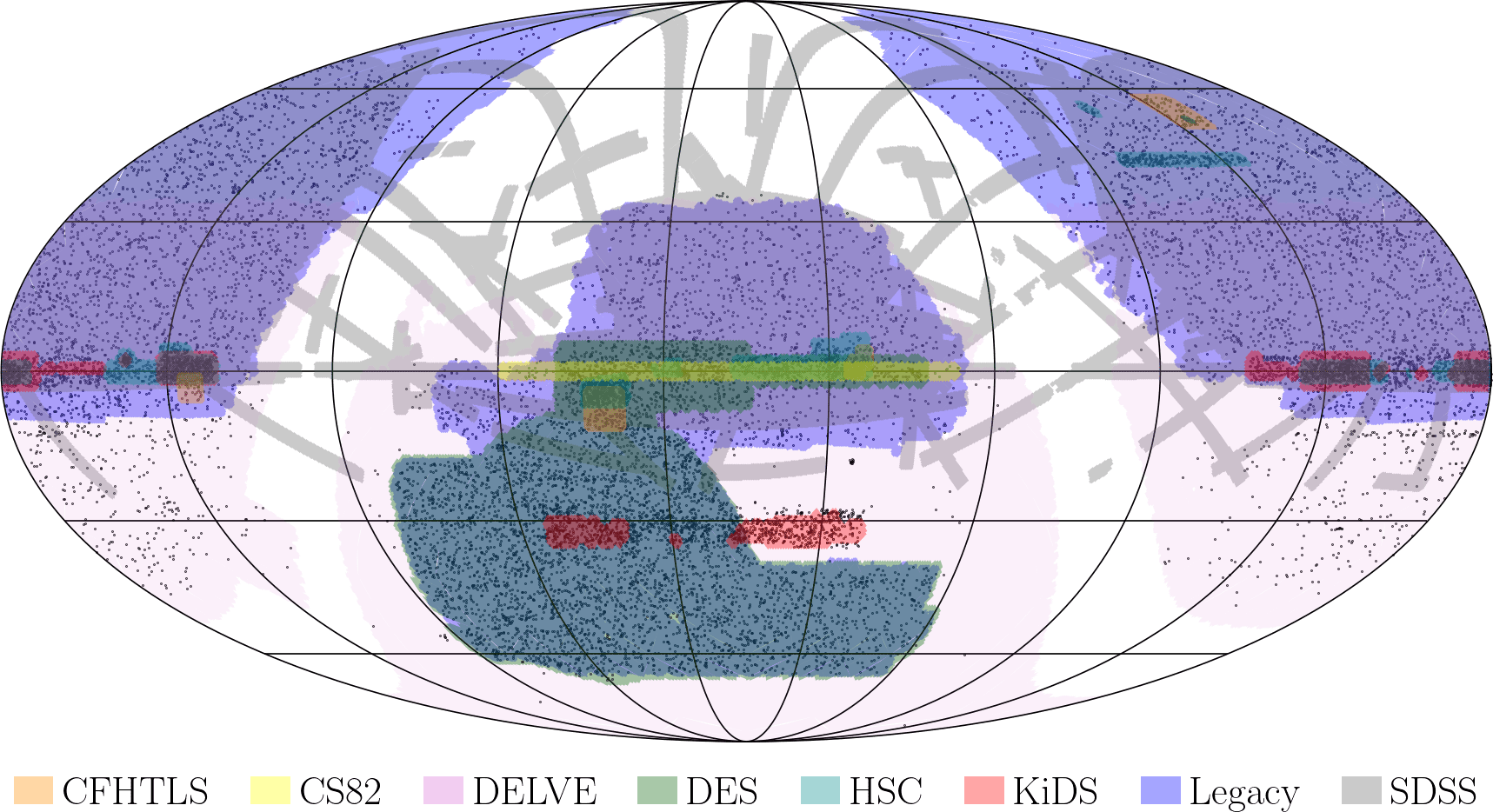}
	\caption{Footprints of the wide-field surveys that are used as sources for matching the photometric information of the SL systems. Each point represents one system from the SL compilation described in Section~\ref{subsec:slfeatures}.}
	\label{fig:photometric-surveys}
\end{figure}

After collecting a list of pre-matched objects, we use the \texttt{xmatch}~\citep{xmatch} tool to perform a uniform cross-match of all gathered catalogs. We also added the complete \ac{CS82} photometric catalog to match objects with \texttt{xmatch}. The cross-matching of all systems in the literature and the surveys mentioned above resulted in 17,154 matched objects, each possessing at least one photometric magnitude, as illustrated in Figure~\ref{fig:photometric-surveys}, overlayed on the respective survey footprint. The photometric cross-match catalog contains magnitudes in the following bands: $u$, $g$, $r$, $i$, $z$, and $y$, as well as photometric redshifts, structural parameters, such as de Vaucouleurs, Petrosian, and exponential radii, information of the data provenance, indicating the survey references and including tile/brick numbers, photometric flags, and internal survey identification.

\subsection{Spectroscopic Catalogs} \label{subsec:spectmatches}

\begin{figure}
	\begin{centering}
		\includegraphics[width=\columnwidth]{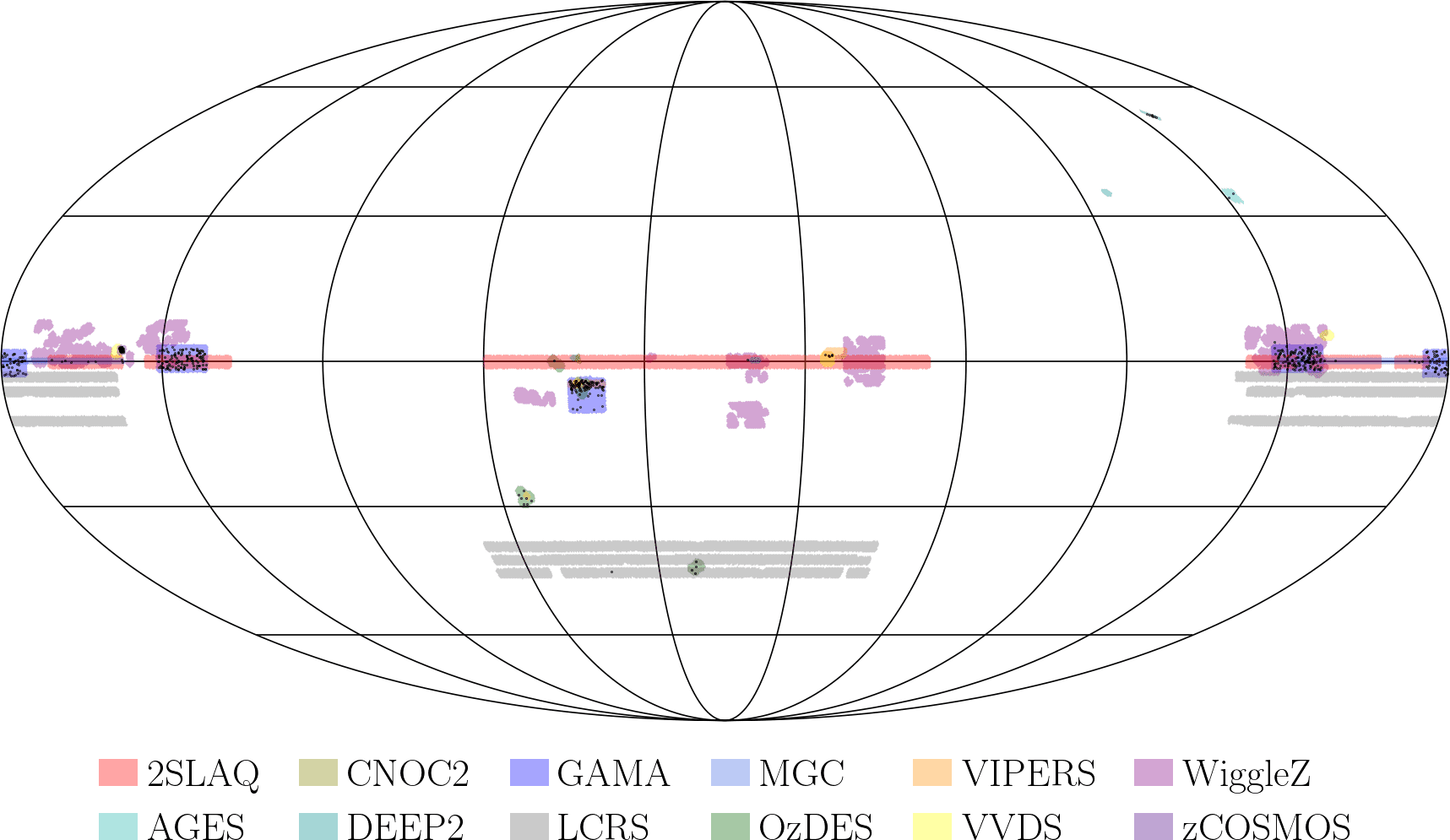}
		\par\end{centering}
	\vspace{0.15cm}
	\begin{centering}
		\includegraphics[width=\columnwidth]{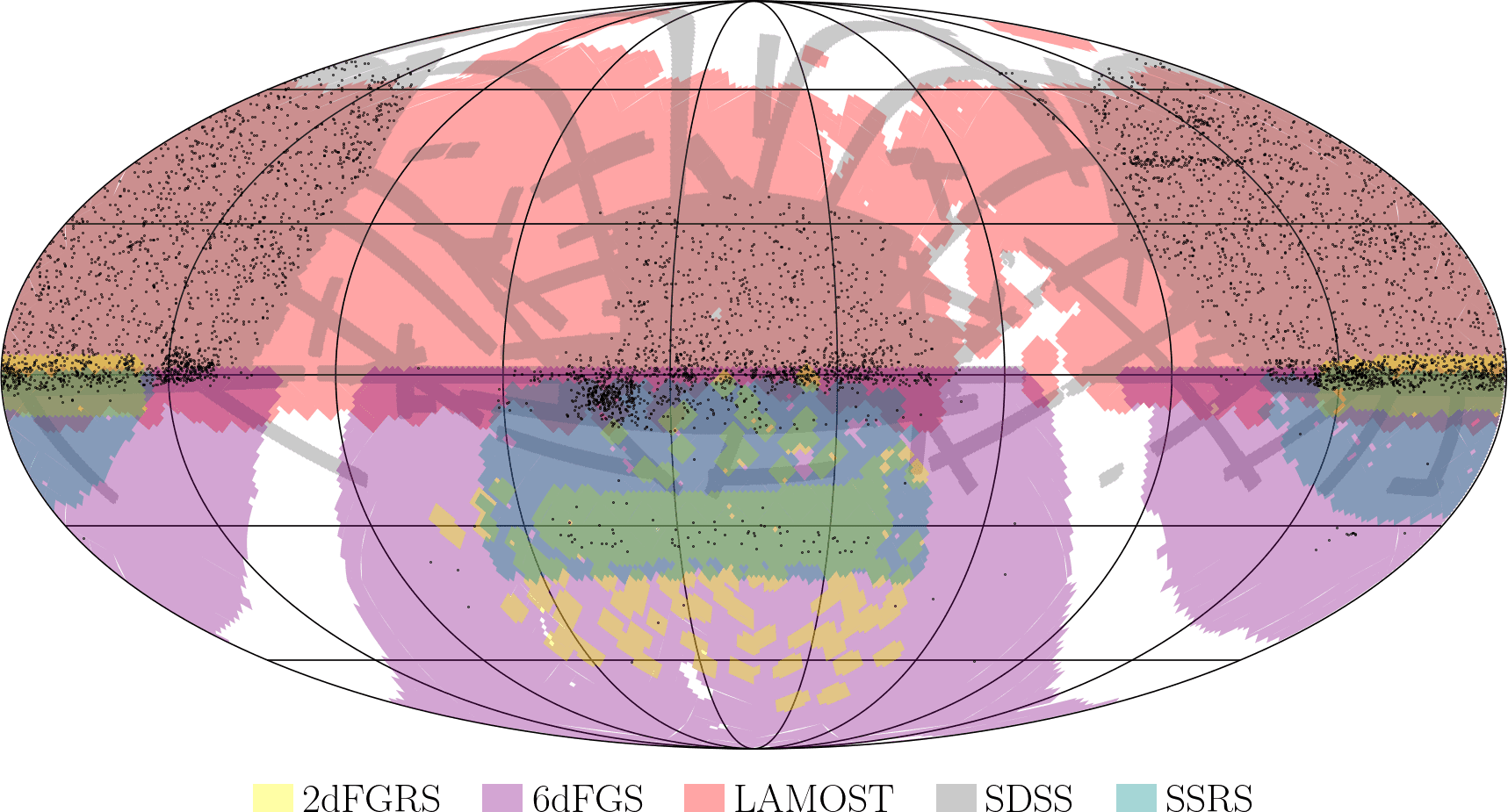}
		\par\end{centering}
	\vspace{0.15cm}
	\begin{centering}
		\includegraphics[width=\columnwidth]{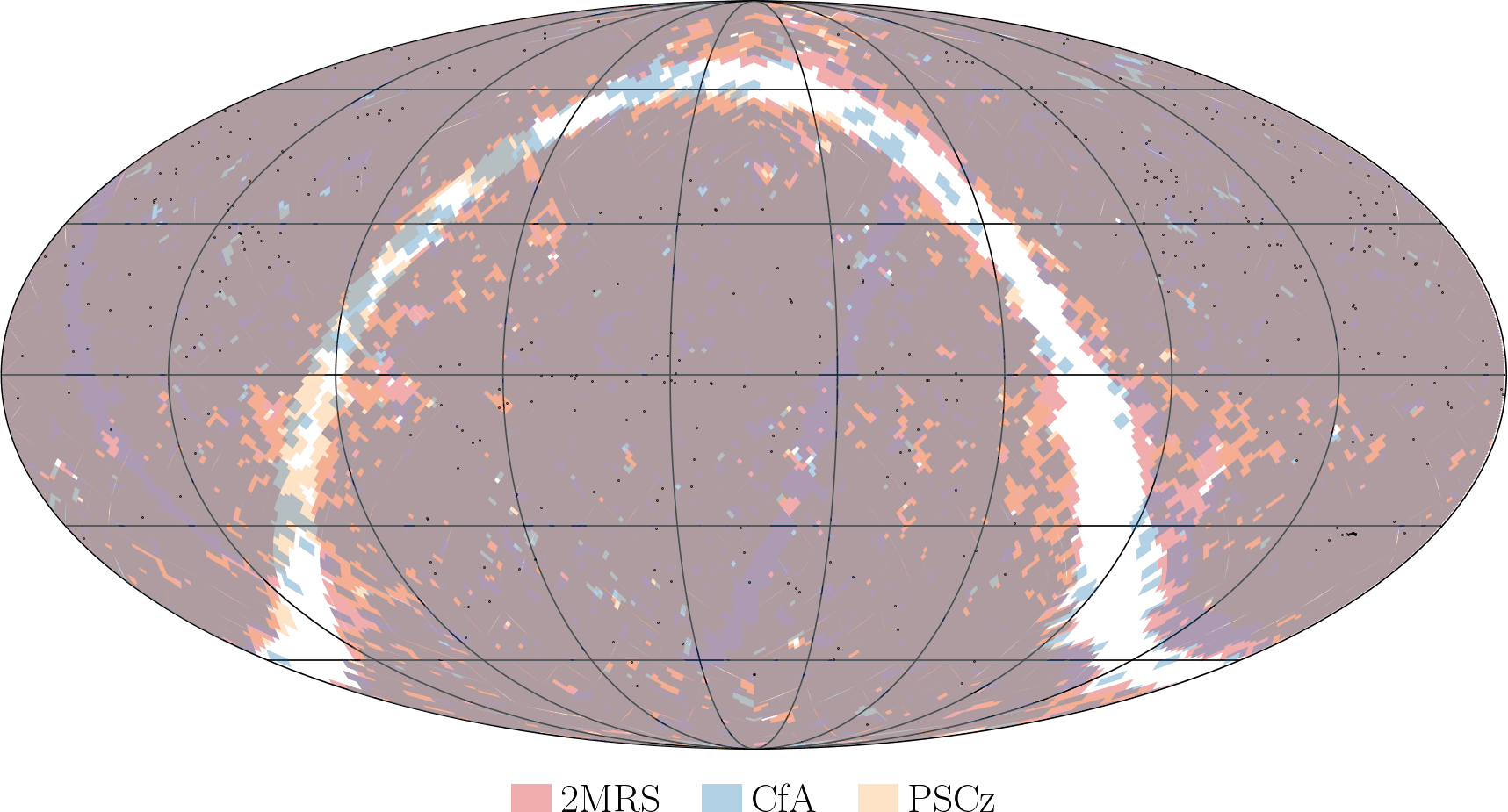}
		\par\end{centering}
	\caption{Footprints of all spectroscopic surveys used to match objects from the literature are categorized into small (top), medium (mid), and large (bottom) areas. The dots represent our matched systems, overlaid on the respective survey footprints.}\label{fig:spectroscopicsurveys}
\end{figure}

The motivation for using spectroscopic data is twofold. First, spectroscopy provides highly precise redshifts for the lensing galaxies, which are fundamental for accurate \ac{SL} modeling. Second, many of these surveys also provide stellar velocity dispersion measurements ($\sigma_v$), an important parameter for lens mass models that is often obtained without additional observational cost. The availability of spectroscopic information, even serendipitously, greatly enhances the scientific value of any \ac{SL} sample.

To build a comprehensive spectroscopic sample, we compiled a list of publicly available surveys with at least 5,000 observed redshifts or coverage pinpoint area of at least 0.5 square degrees~\citep{Surveys} to cross-match with our \ac{SL} catalog. The footprints of these surveys are shown in Figure~\ref{fig:spectroscopicsurveys}, where they are categorized by their area coverage. Following this structure, we list them below:

\begin{itemize}
    \item Small-area ("pencil-beam") surveys: These deep surveys provide valuable data in very specific fields, including the final data release of the \ac{2SLAQ}~\citep{2SLAQ}, the \ac{CNOC2}~\citep{CNOC2}, the third data release of the \ac{GAMA} (``SpecObj'' table from ``SpecCat v27'' catalog in~\citet{GAMA}), the \ac{MGC}~\citep{MGC}, the final data release of \ac{VIPERS}~\citep{VIPERS}, the first data release of the \ac{WiggleZ}~\citep{WiggleZ}, the \ac{AGES}~\citep{AGES2}, the forth data release of \ac{DEEP2}~\citep{DEEP2}, the \ac{LCRS}~\citep{LCRS} surveys, the second release of the \ac{OzDES} (data available at Data Central\footnote{\url{https://datacentral.org.au}};~\citet{OzDES}), the final data release of \ac{VVDS}~\citep{VIMOS}, and the \ac{zCOSMOS}~\citep{zCOSMOS}.
    \item Medium-area surveys: These surveys focus on smaller, targeted regions and include the final release of the \ac{2dFGRS} (``best spectroscopic observations'' table from~\citet{2dF}), the third data release of the \ac{6dFGS}~\citep{6dF}, the value-added catalog ``Central Velocity Dispersion Catalog of LAMOST-DR7 Galaxies'' of \ac{LAMOST}~\citep{LAMOSTVAC}, the seventeenth release of the \ac{SDSS} (table ``specobjall'' from Astro Data Lab service), and the \ac{SSRS}~\citep{1998AJ....116....1D}.
    \item Large-area surveys: Covering vast portions of the sky, these surveys include the \ac{2MRS}~\citep{2MASS}, \ac{CfA}~\citep{CfA} and \ac{PSCz}~\citep{2000MNRAS.317...55S}.
\end{itemize}

We have also included two value-added catalogs containing spectroscopic redshifts based on matches of objects from photometric surveys and public spectroscopic surveys. The first value-added catalog comes from the third data release of the \ac{HSC} catalogs, using the table ``specz'' from matches of objects from their Wide and Deep/Ultra fields. The second value-added catalog containing spectroscopic redshifts comes from the column ``zspec'' from the ``photoz'' table from the second data release of the \ac{DELVE} survey (the same table used to add photometric redshifts for this survey in the previous Section).

Before cross-matching, we standardized the compiled catalogs through a data cleaning process. This involved two main steps. First, we performed preliminary cleaning by removing entries with invalid data, such as NaN values or non-physical measurements (e.g., negative measurements), and by converting the coordinates of older surveys (\ac{CfA}, \ac{CNOC2}, \ac{LCRS}, \ac{PSCz}, and \ac{SSRS}) to the J2000 reference frame.

Second, to ensure the reliability of the redshift measurements, we retained only the spectra with the highest confidence flags, following the specific quality criteria recommended by each survey collaboration. The flags used for the main surveys were:

\begin{itemize}
    \item \ac{2dFGRS}: quality>=3
    \item \ac{2SLAQ}: qz\_z2S = 1
    \item \ac{6dFGS}: QUALITY = 4
    \item \ac{DEEP2}: q\_z = 2, 3, 4
    \item \ac{GAMA}: NQ > 2
    \item \ac{MGC}: q\_z = 3, 4, 5
    \item \ac{OzDES}: Q = 3, 4
    \item \ac{SDSS}: zWarning=zWarning\_noqso=0
    \item \ac{VIPERS}: flag = 2, 3, 4
    \item \ac{VVDS}: flag = 2, 3, 4
    \item \ac{WiggleZ}: q\_z = 3, 4, 5
\end{itemize}

With the cleaned and quality-filtered catalogs, we performed the cross-match using the \texttt{xmatch}~\citep{xmatch} tool to identify matches within a radius of $1^{\prime\prime}$. This process resulted in 6,440 objects from our \ac{SL} sample having a spectroscopic counterpart, as illustrated in Figure~\ref{fig:spectroscopicsurveys}, where the matched systems are overlaid on the survey footprints. Most of the matches are with \ac{SDSS} data, thanks to its extensive area coverage. Given the matching radius of $1^{\prime\prime}$, we only expect matches for the central lens galaxy, which is appropriate for single lens galaxy systems, the primary focus of this work. Searching for spectroscopic counterparts of other group members or lensed images would require dedicated analysis and is left for a future version.

The final spectroscopic cross-match catalog includes redshifts, velocity dispersions, magnitudes in multiple bands, and the original spectra quality flags. It also contains essential metadata for provenance, such as the original survey name, object identifiers (e.g., Plate-MJD-Fiber for \ac{SDSS}), and observation dates, when available.

\section{Consolidated Catalog} \label{sec:consolidated}

As described in the previous sections, we constructed three main catalogs: a literature compilation and two cross-matched catalogs with data from photometric and spectroscopic surveys (see Section~\ref{subsec:slfeatures}  and Sections~\ref{subsec:photomatches} to ~\ref{subsec:spectmatches} for details). The compilation process often resulted in multiple data entries for a single physical parameter of a given system, which is identified by the unique key JNAME. This is known as a one-to-many relationship: one system (JNAME) corresponds to many potential values for features like redshift or magnitude. To create a directly usable dataset, we developed a consolidated catalog that resolves these multiple entries into a single value for each feature. This section details the hierarchical rules and logic applied to generate this final one-to-one catalog.

The catalogs already feature a few columns with unique values, including JNAME as the system identifier and coordinates (RA and Dec) derived from JNAME. Another feature with a single one-to-one value is the System Type (``Single Lens Galaxy'', ``Non Single Lens Galaxy'', and ``Non Single Lens Galaxy Merged''). The challenge of creating a consolidated catalog is assigning unique values to columns such as lens and source types, Einstein radius, magnitudes, redshifts, and velocity dispersion.

We assigned a unique value to lens/source types based on consensus among the original references. If all sources agreed on the classification (e.g., all listed ``galaxy''), we adopted that value. However, in cases of disagreement (e.g., one reference classified the object as a ``galaxy'' and another as a ``cluster''), the corresponding entry was left blank to avoid introducing ambiguity.

To consolidate the Einstein radius ($\theta_E$), we established a two-step process. First, we filtered the values based on the estimation method. Since the accuracy of $\theta_E$ is used in many applications, we only considered values derived from fitting parametric lens models—such as the \ac{SIS}, \ac{SIE}, or more general power-law profiles. We explicitly excluded simpler estimates, such as those based on half the image separation. For instance, values from studies like~\citet{Nord_et_al._2019,Amvrosiadis_et_al._2018,Kubo_et_al._2008} were not included if they were not derived from lens modeling.

If multiple model-derived values remained for a single system, we applied a ``Reference Hierarchy''. This is a simple prioritization scheme where references are sorted by publication date, giving precedence to the most recent parameter estimation or measurement.

To assign a single magnitude value for each of the ($u$, $g$, $r$, $i$, $z$, $y$ bands, we established a prioritized selection procedure based on the data source:

\begin{enumerate}
    \item ``Photometric Hierarchy'': Priority was given to our photometric cross-matches (Section~\ref{subsec:photomatches}), with surveys ranked by the primary mirror size of the telescope (Section~\ref{sec:cutouts}). This choice is motivated by the fact that larger apertures generally yield more reliable photometry.
    \item ``Spectroscopic Hierarchy'': If a system had no counterpart in the photometric catalogs, we used magnitudes from the spectroscopic cross-matches (Section~\ref{subsec:spectmatches}). These surveys were ranked by their total number of cross-identifications in our sample, prioritizing the most comprehensive source.
    \item ``Reference Hierarchy'': For systems without a counterpart in any cross-match, we reverted to the magnitude values from their original source catalog, applying the date-based ``Reference Hierarchy'' described previously.
\end{enumerate}

Assigning unique source and lens redshifts ($z_S$, $z_L$) required an hierarchical approach. For source redshifts, which are primarily available in the original \ac{SL} catalogs, we applied the ``Reference Hierarchy'', prioritizing spectroscopic over photometric measurements when both were available from different sources.

\begin{figure}
	\includegraphics[width=\columnwidth]{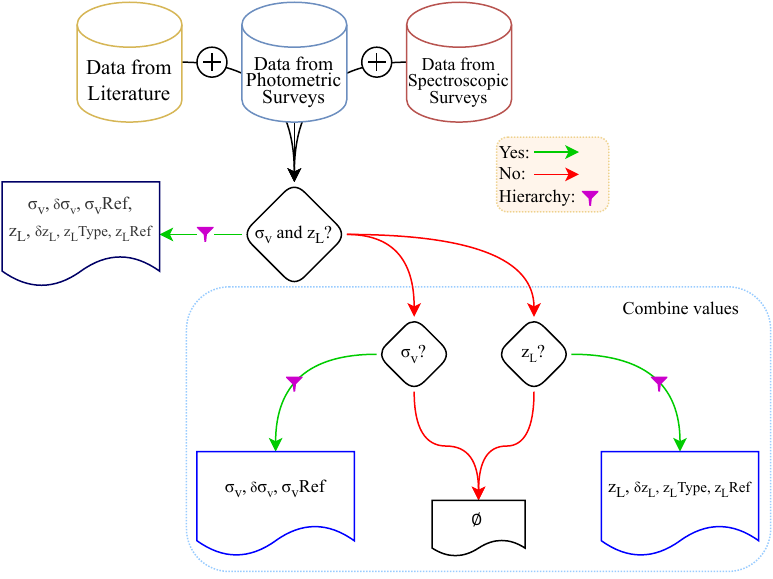}
	\caption{Workflow for assigning a single, adopted velocity dispersion ($\sigma_v$) and lens redshift ($z_L$). The process follows a clear hierarchy, prioritizing spectroscopic data from our cross-matches and literature compilation over photometric redshifts.}
	\label{fig:diagramredshift}
\end{figure}

For the lens redshift, our primary goal was to obtain the most precise value available. Therefore, we prioritized spectroscopic redshifts ($z_{spec}$) over photometric ones ($z_{phot}$) due to their much higher accuracy. The workflow for assigning the final adopted lens redshift, also illustrated in Figure~\ref{fig:diagramredshift}, was as follows:

\begin{enumerate}
    \item First, we checked for a spectroscopic redshift from our cross-matched survey catalogs.
    \item If none was found, we used a spectroscopic redshift from our literature compilation. This step also allowed us to incorporate velocity dispersion measurements, which are intrinsically spectroscopic.
    \item Only if no spectroscopic redshift was available from any source, we adopted a photometric redshift.
\end{enumerate}

Throughout this process, we kept track of the origin and nature (spectroscopic or photometric) of the final adopted redshift for provenance.

In this Section, we have detailed the procedures to create a unified catalog with a single, adopted value for each key parameter per system. To ensure reproducibility and facilitate backtracking, we have provided provenance for each consolidated column. In Section~\ref{sec:lastberu-applications}, we highlight potential scientific applications enabled by this unified dataset and provide details on its public availability.

\section{Images from Wide-Field Surveys} \label{sec:cutouts}

Images of gravitational lensing systems serve several purposes, such as fitting a lens and source model based on a specific gravitational potential (see, for example,~\citet{2018PDU....22..189B} and~\citet{pyautolens}) or for the identification and simulation of these systems~\citep{2022arXiv220309536B}. This work aims to prepare a dataset to develop and test infrastructure for the upcoming Rubin \ac{LSST} data. Therefore, a core component of this work is a comprehensive collection of image cutouts for the \ac{SL} candidate systems in our compilation, gathered from multiple ground-based surveys and bands. We argue this is the most extensive collection possible with current publicly available data, making the LaStBeRu database an essential ground-truth sample for developing and testing algorithms for the next generation of wide-field surveys, such as the \ac{LSST}.

\begin{figure}
	\includegraphics[width=\columnwidth]{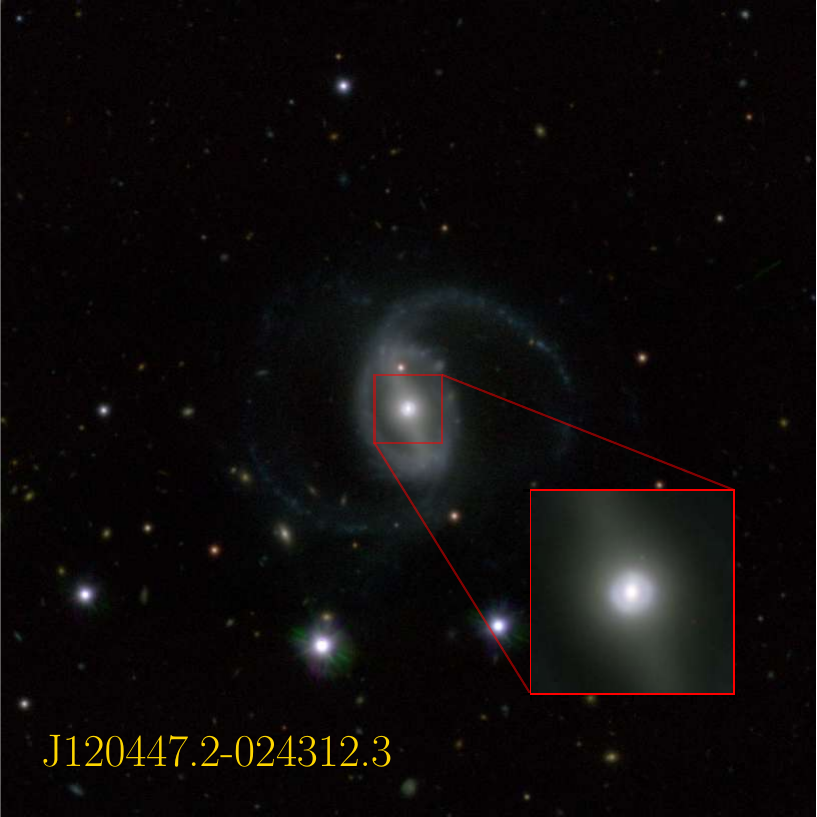}
	\caption{An example of the same object at different cutout scales: the galaxy J120447.2-024312.3 from the \acs{LinKS} sample~\citep{LinKS} seen in the KiDS imaging data.}
	\label{fig:J12044}
\end{figure}

We chose to provide cutouts in two sizes for all imaging data: $20^{\prime\prime}$ and $4^\prime$. The first ($20^{\prime\prime}$) is a standard size that typically encompasses galaxy-galaxy scale systems and has been commonly used for \ac{SL} searches, especially using \ac{ML} techniques, such as \ac{CNN}s, which require images of a fixed size (e.g.,~\citet{2019A&A...625A.119M,2022arXiv220309536B}). The second image size ($4^\prime$) was chosen for several reasons: i) to encompass the largest \ac{SL} systems, namely massive clusters; ii) to allow us to evaluate the environment of the lens; and iii) to help distinguish bona fide lenses from contaminants, such as the inner rings of large, nearby (low-redshift) late-type galaxies, which can be mistaken for Einstein rings in smaller cutouts. For the first motivation, we considered the prediction for the largest Einstein radius of the Universe~\citep{10.1111/j.1365-2966.2008.14154.x}, which should be approximately $\sim100^{\prime\prime}$, and the observation of the largest known lensed image by a cluster~\citep{Zitrin_2009}, whose distance from the lens center is $\sim150^{\prime\prime}$. Thus, we decided to use cutouts of $4^{\prime} = 240^{\prime\prime}$, such that all strongly lensed images would fit in such cutouts. This format also encompasses the typical size of group and cluster scale lenses, so that it is enough to evaluate the environment to characterize the lens and to look for nearby galaxy overdensities. As illustrated in Figure~\ref{fig:J12044}, a visual inspection of many systems at this scale is sufficient to identify if a putative Einstein ring is actually an inner ring of a late-type galaxy. This system was classified as a \ac{SL} in a search carried out with \ac{CNN} using small cutouts on \ac{KiDS} data~\citep{LinKS} (as in the figure inset). Looking at the larger cutout, it becomes evident that the ring feature is part of a late-type galaxy and is not a bona fide lensed source.

We provide all cutouts using the pixel scale of the original images, which differ due to the telescope optics and intrinsic pixel size. Therefore, while all cutouts from the different surveys have approximately the same angular size, the number of pixels in the images may vary.

The main properties of the surveys from which we extracted the cutouts are summarized below. We note that specific data release versions are provided for reproducibility.

The Canada-France-Hawaii Telescope Legacy Survey~\citep{CFHTLS} was carried out with the MegaCam instrument at the 3.58 m \ac{CFHT}. The survey collected approximately 155 square degrees of data in the $u$, $g$, $r$, $i$, and $z$ bands. We use the images reprocessed by the \ac{CFHTLenS} consortium~\citep{CFHTLenS}, which optimized the data reduction for weak lensing studies. The \ac{CS82} Survey~\citep{CS82} covered approximately 173 $\deg^2$ in the $i$ band, also with MegaCam, and employed the \ac{CFHTLenS} processing pipeline. Data from the Red-sequence Cluster Survey 2~\citep{2011AJ....141...94G}, also conducted with MegaCam, was processed by the \ac{RCSLenS} project~\citep{RCSLenS} using the same pipeline as \ac{CFHTLenS}, covering an area of 785 $\deg^2$ in the $g$, $r$, $i$, and $z$ bands.

The \ac{KiDS}~\citep{KiDS} (Data Release 4) collected approximately 1,350 square degrees of imaging data in bands $u$, $g$, $r$, and $i$, with the OmegaCAM camera on the 2.61 m \ac{VST}. The \ac{HSCSSP}~\citep{HSC} (Data Release 3) covered $\sim$1430 $\deg^2$ in the $g$, $r$, $i$, $z$, and $y$ bands using the Hyper Suprime-Cam on the 8.2 m Subaru telescope. The \ac{DES}~\citep{2005astro.ph.10346T} (Data Release 2) covered an area of over 5,000 $\deg^2$ in the $g$, $r$, $i$, $z$, and $y$ bands with the \ac{DECam} on the 4 m Blanco telescope.

The DESI \ac{Legacy}~\citep{Legacy} (DESI \ac{Legacy} for short) (Data Release 9 and 10) combined data from three distinct projects to cover over 20,000 $\deg^2$. It used the Blanco 4 m telescope with \ac{DECam} for the \ac{DECaLS}, the Bok 2.3 m telescope with the 90Prime camera for the \ac{BASS}, and the Mayall 4 m telescope with the Mosaic-3 camera for the \ac{MzLS}, collecting data in the $g$, $r$, and $z$ bands.

\subsection{Collecting Image Cutouts} \label{subsec:imageprocess}

We collected image cutouts, which are small portions of larger survey images centered on the coordinates of our targets, at two different angular scales. The smaller cutouts, with a side of $20^{\prime\prime}$, are designed for detailed analysis of galaxy-scale lens systems. The larger $4^{\prime}$ cutouts allow for the study of the systems' environment, such as identifying group or cluster members. Compiling these cutouts from all aforementioned imaging surveys in the pre-defined sizes is not a trivial task, since each project has their own form for retrieving the data.\footnote{It is worth noting that some \ac{SL} candidates may not be visible in these ground-based images. This can occur if the lensed features are too faint for the survey's depth, if the system is too compact to be resolved given the atmospheric seeing, or if the candidate was originally selected at other wavelengths (e.g., sub-millimeter).}

For projects such as \ac{CS82} and \ac{KiDS}, which do not have an automated image retrieval interface, we produced the cutouts locally from their full sets of processed images. To do this, we used the World Coordinate System (WCS) information in the \ac{FITS} headers to create smaller images centered on our targets using the Astropy library~\citep{astropy}. The final processed \ac{CS82} data is stored in the same system we used to generate the LaStBeRu database. In the case of \ac{KiDS}, we downloaded the full image tiles from their fourth data release~\citep{KiDSDR4}, including also tiles from previous releases that were not part of this release.

\ac{CFHTLenS}\footnote{\url{https://www.cfhtlens.org}} and \ac{RCSLenS}\footnote{\url{https://rcslens.org}} have cutout services, which were used to retrieve cutouts of 108 and 1292 pixels on a side, corresponding to the $20^{\prime\prime}$ and $4^{\prime}$ scales, respectively.

In the case of \ac{HSC} we used the DAS Cutout tool\footnote{\url{https://hsc-release.mtk.nao.ac.jp}} to collect images from their  Data Release 3~\citep{HSCDR3} (for the Wide, Deep, and Ultradeep layers).

To obtain data from the DESI \ac{Legacy} and \ac{DES} surveys, we used the Astro Data Lab \ac{SIA} service\footnote{\url{https://datalab.noirlab.edu}}. Specifically for the DESI \ac{Legacy}, we used Data Release 8, since missing tiles were reported in the $z-$band for some systems in Data Release 9. Finally, we used Data Release 2 (DR2) for images from the \ac{DES}~\citep{DESDR2} survey.

After acquiring the image cutouts, we cleaned the sample through a multi-stage process. First, we addressed cases where a single system from the LaStBeRu catalog was observed in multiple cutouts within the same survey. To select the single best image, we used information from the \ac{FITS} header, as detailed in Table~\ref{tab:cutouts-info}. Our selection prioritized the best seeing, followed by the longest exposure time. For the \ac{Legacy} survey, which lacks this header information, we instead selected the tile with the lowest \ac{PSF} depth. Similarly, for \ac{DES} images, we chose the tile with the highest object count.

\begin{table}
    	\centering
    	\caption{Total and unique cutouts collected for each imaging survey are presented, including header information used to select each image in cases where multiple images of the same object per survey and band were available. The numbers on the second column correspond to 20 arc-second cutouts, while those on the third column denote 4 arc-minute cutouts.}
    	\label{tab:cutouts-info}
    \resizebox{\columnwidth}{!}{
    	\begin{tabular}{cccc}
        \hline
        Survey & Total & Unique & Header Info.\\
        \hline
        \ac{CFHTLenS} & 4262/4262 & 853/853 & \textsf{SEEING},\textsf{TEXPTIME}\\
        \ac{CS82} & 870/850 & 870/850 & \textsf{SEEING},\textsf{TEXPTIME}\\
        \ac{DES} & 43688/43688 & 8913/8913 & \textsf{count}\\
        \ac{HSC} & 33975/33898 & 7568/7552 & \textsf{HIERARCH\_variance\_scale}\\
        \ac{KiDS} & 14946/13977 & 4459/4311 & \textsf{PSF\_FWHM}\\
        \ac{Legacy} & 81034/81017 & 28043/28043 & \textsf{psfdepth\_g,r,z}\\
        \ac{RCSLenS} & 5219/5219 & 1465/1465 & \textsf{SEEING},\textsf{TEXPTIME}\\
        \hline
    	\end{tabular}
    }
\end{table}

Next, we removed cutouts where no objects could be identified. An initial check revealed that some files consisted entirely of zero-value arrays, particularly affecting approximately 11\% of objects in \ac{CFHTLenS} and 18\% in \ac{RCSLenS}. This issue appears more prevalent in surveys using the MegaCam instrument, suggesting a potential problem in their data retrieval pipeline. To systematically identify and discard all cutouts lacking detectable sources, we used the \ac{SEP} library~\citep{1996A&AS..117..393B,Barbary2016} to generate a source catalog for each $20^{\prime\prime}$
 image. If \ac{SEP} detected no sources, the cutout was considered unsuitable for fitting or visual inspection. This method revealed that around 16\% of \ac{KiDS} images were empty; for other surveys, the detection failure rate was less than 6\%.

\begin{figure}
	\centering
	\includegraphics[width=\columnwidth]{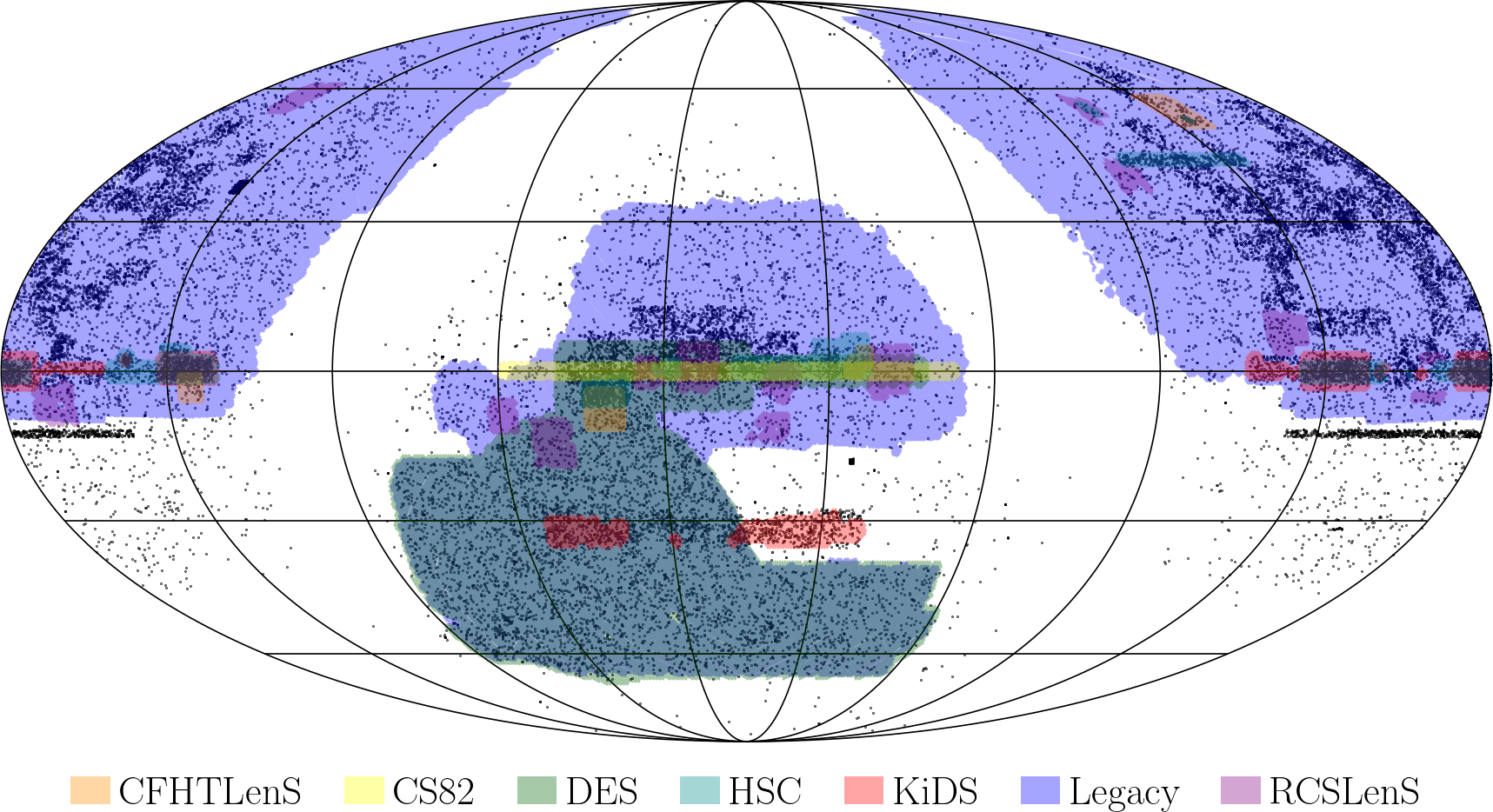}
    \caption{Mollweide projection featuring catalog objects (dots) and survey footprints for cutout acquisition. Dots outside the footprints do not have any cutouts available.}
	\label{fig:cutouts-footprints}
\end{figure}

Finally, because \ac{SL} candidate systems can be on the edge of the imaging bricks, we implemented an image size filter, discarding $20^{\prime\prime}$
 cutouts if their height or length deviated by more than a $1^{\prime\prime}$ tolerance. This test affected only about 4\% and 2\% of \ac{Legacy} and \ac{DES} images, respectively. Figure~\ref{fig:cutouts-footprints} shows the final catalog objects with at least one valid image cutout, overlaid on the survey footprints.

\begin{figure}
	\centering
	\includegraphics[width=\columnwidth]{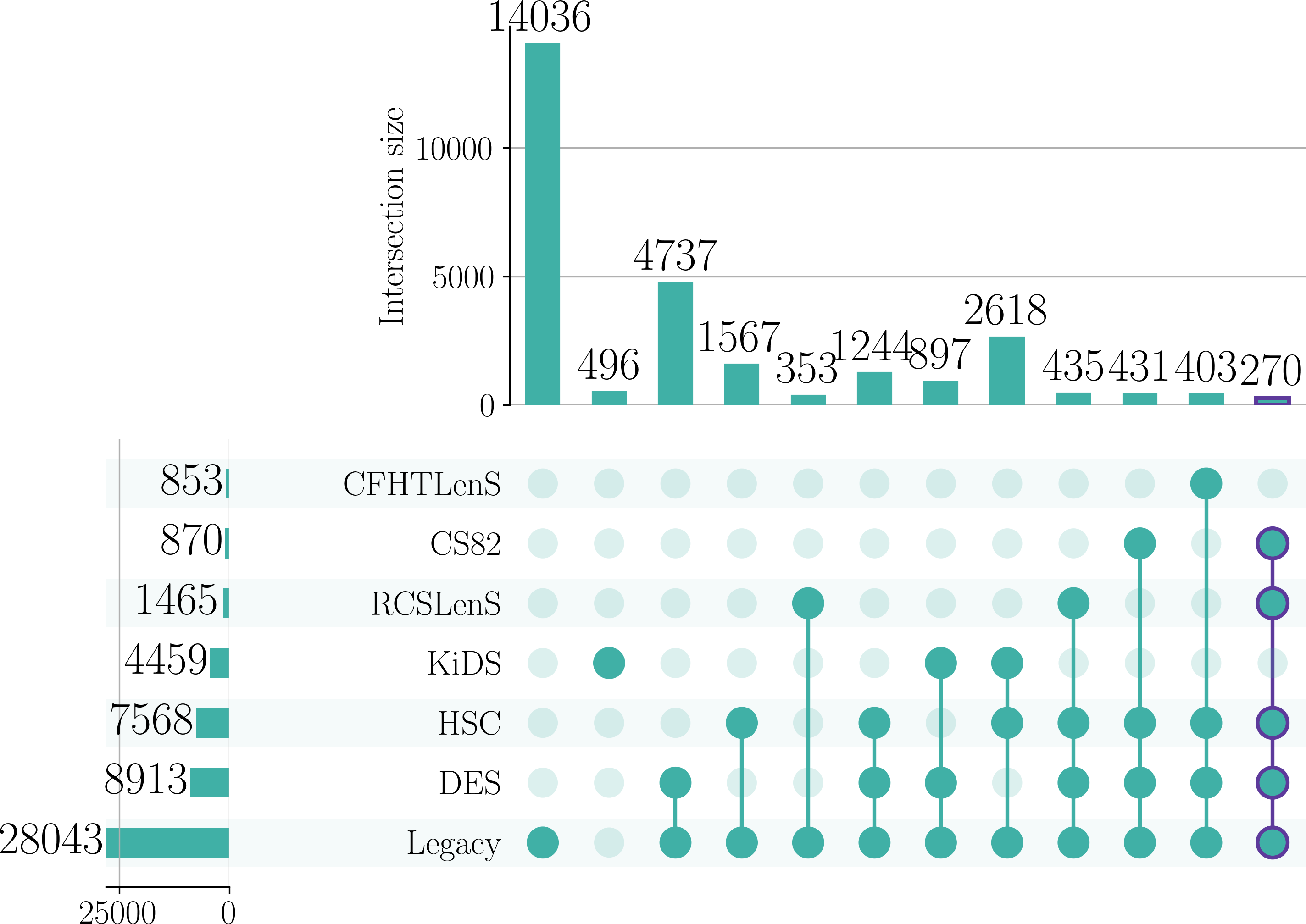}
	\caption{Intersections of unique systems with cutouts of $20^{\prime\prime}$ are available in the seven imaging surveys included in this work. The numbers on the bottom left panel show the number of unique objects with cutouts in each survey. The lines indicate the intersection of the surveys, and the UpSet plot on the top shows the number of systems with cutouts in the intersecting surveys. A dot without any line corresponds to the number of systems that only have imaging on that specific survey. For simplicity, we show only intersections and matches with at least 250 systems.}
	\label{fig:cutouts-intersections}
\end{figure}

As there is a large overlap between the footprints of the surveys considered, many systems will have cutouts in more than one survey. This is useful, for example, to assess the impact of the survey properties, such as depth and \ac{PSF} on the detection and modeling of the systems. In Figure~\ref{fig:cutouts-intersections}, we show the number of cutouts that intersect among the surveys considered. As expected, since the DESI Legacy Survey has the most extensive coverage, it contains the largest number of overlapping systems among several surveys and has the largest number of unique systems (29,068). The two rightmost columns in this Figure show the number of systems with the most significant overlap among surveys: 270 systems with cutouts available in \ac{CS82}, \ac{RCSLenS}, \ac{HSC}, \ac{DES}, and \ac{Legacy} and 46 systems with cutouts available in \ac{CFHTLenS}, \ac{CS82}, \ac{HSC}, \ac{DES}, and \ac{Legacy} surveys (omitted in the Figure).

\begin{figure}
	\begin{centering}
		\setkeys{Gin}{height=0.33\columnwidth}
		\includegraphics{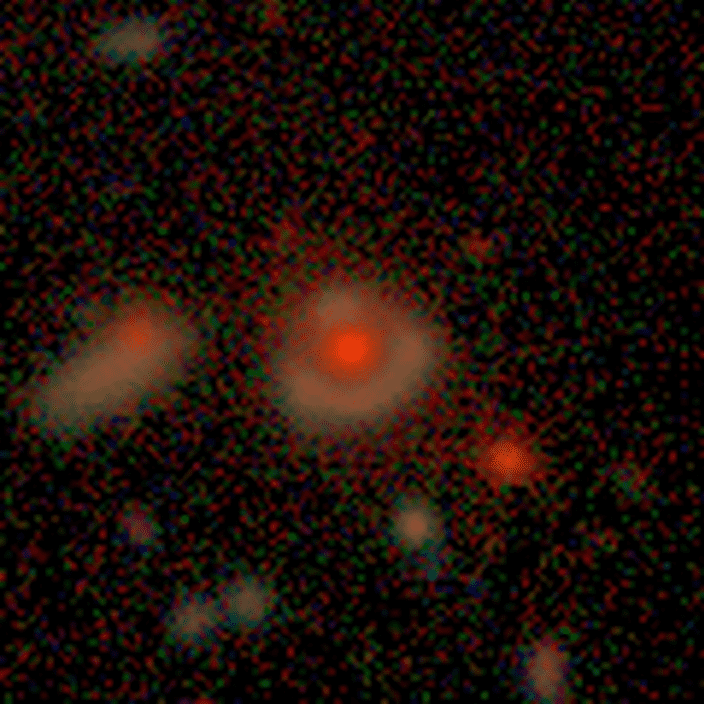}
		\includegraphics{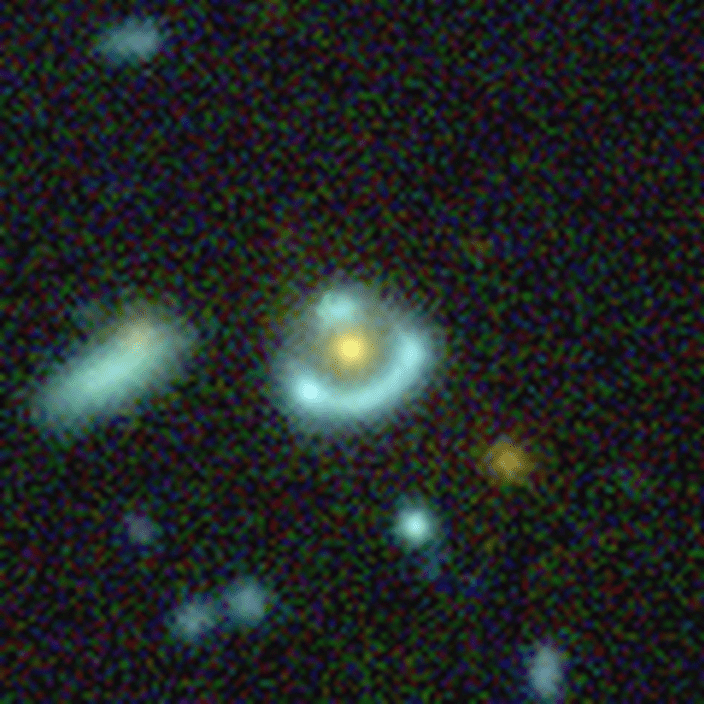}
		\includegraphics{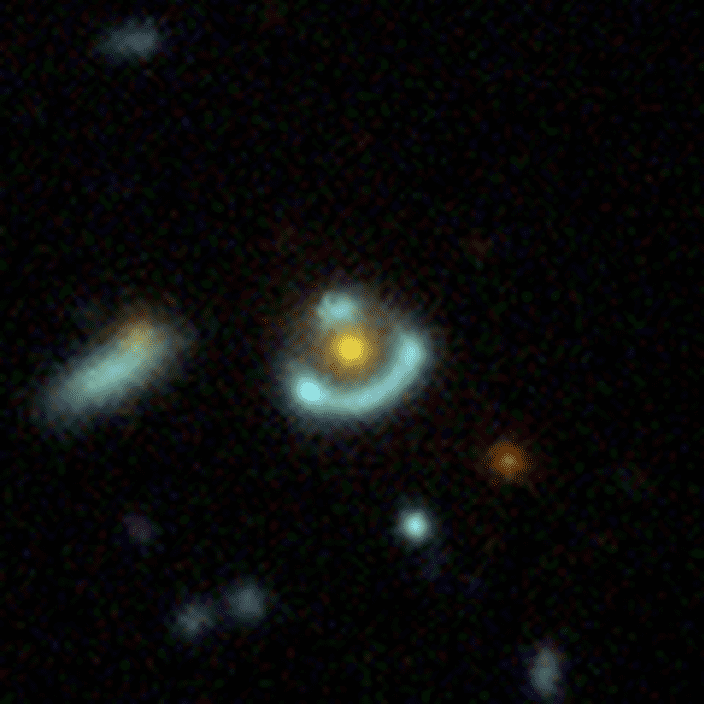}
	\end{centering}
	\caption{Colored images of J010238.3+015856.8 from \acs{HSC} images using \texttt{make\_lupton\_rgb}, STIFF, and Trilogy (from left to right).} \label{fig:compmethods}
\end{figure}

\begin{figure*}
	\begin{centering}
		\includegraphics[width = 150mm]{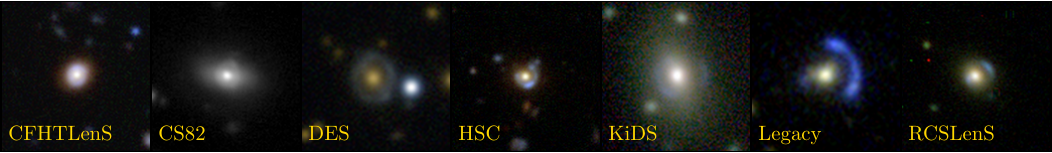}
		\par\end{centering}
	\begin{centering}
		\includegraphics[width = 150mm]{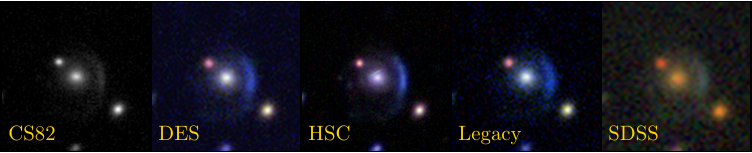}
		\par\end{centering}
    \caption{Top: Examples of different strong lensing systems in each survey. Bottom: system J015750.4$-$003818.5 seen in various surveys.}
    \label{fig:Mosaics}
\end{figure*}

\subsection{Image Combination} \label{subsec:combination}

Combining astronomical monochromatic (single-band) images into color images (usually composed of red, green, and blue) is a tricky task because this process involves, for example, calibration of color saturation, gain, and image normalization. Fortunately, several tools make this process easy, such as the \texttt{make\_lupton\_rgb} module from Astropy, STIFF~\citep{STIFF}, and Trilogy~\citep{Trilogy}. The main problem is setting optimal parameters for visualization. In Figure~\ref{fig:compmethods}, we present a comparison between the three methods mentioned above for one of the systems in our compilation.

We performed experiments using different lensing systems to estimate the best global parameters for each image combination technique that works best for the surveys we used in the catalog (these parameters are available in the repository mentioned in the Data Availability Section). The Astropy module uses the method proposed in~\citep{Lupton}, tends to make reddish images. However, STIFF and Trilogy tend to produce bluish images. The images processed with STIFF and Trilogy were less noisy than those of the Astropy module. We emphasize that we used the same parameters for all tested images, and tweaking globally for processing multiple images might produce unstable processed images.

\begin{figure}
	\centering
	\includegraphics[width=\columnwidth]{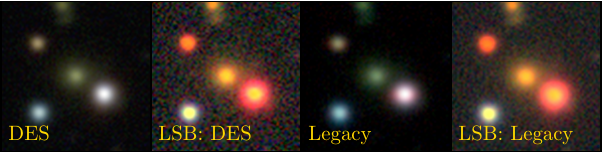}
	\caption{Example of artifacts of the object J213822.0+011648.9 observed in two surveys with Trilogy combined images and LSB images.} \label{fig:fakerings}
\end{figure}

For each of the images obtained in the previous Section, we have processed them using Trilogy\footnote{\url{https://pypi.org/project/trilogy}} software. The galaxy shown in Figure~\ref{fig:J12044} is an example of an image processed with Trilogy. This software efficiently combines images from 3 bands (R: bands $z$, $y$, or $i$; G: band $r$; B: band $g$) without saturating the image brightness. In this way, we provide \ac{RGB} and single-band (black-and-white) images. In Figure~\ref{fig:Mosaics}, we present examples of cutouts from all surveys used. In the first row, we showed many lenses, most of them arcs and Einstein rings, and systems classified as ``Single Lens System'': J085856.0-010208.4 (\ac{CFHTLenS}), J234111.6+000018.7 (\ac{CS82}), J000644.6-442950.5 (\ac{DES}), J140929.7-011410.8 (\ac{HSC}), J221339.0-331156.3 (\ac{KiDS}), J120602.1+514229.4 (\ac{Legacy}), J213758.0-012923.9 (\ac{RCSLenS}). In the second row, the variations of each survey, such as exposure time and observational seeing, allow us to observe additional properties within the same system.

Surveys also provide combined color images using some internal methods. The \ac{LSB} has compiled composite color cutouts for many surveys, including their own (\ac{Legacy} Surveys (\ac{DECaLS}, \ac{MzLS}, \ac{BASS}, \ac{DECaPS}, and \ac{unWISE})) and other images from \ac{SDSS}, \ac{DES}, and \ac{HSC}. In addition to the processed images obtained with Trilogy, we have added processed cutouts from the \ac{LSB}\footnote{\url{https://www.legacysurvey.org/viewer}} tool, including \ac{SDSS} images for completeness. Depending on the parameters used to combine the images, some artifacts, such as the fake rings shown in Figure~\ref{fig:fakerings}, may arise in the \ac{LSB} images. Comparing \ac{LSB} to Trilogy images, the former tends to make objects reddish, like Lupton's method, which could affect the identification of blue arcs and red lenses. However, while the Trilogy images appear less saturated, the \ac{LSB} emphasizes physical features. Therefore, choosing a combination of imaging methods can affect visual inspections and system confirmations.

\section{The slcomp Framework and LaStBeRu Database} \label{sec:lastberu}

The Strong Lens Compilation ``slcomp'' framework is summarized in Figure~\ref{fig:framework}, which is responsible for generating the database. This framework aggregates data for confirmed and candidate \ac{SL} systems, enriches this information through cross-matching with wide-field surveys, and compiles an extensive mosaic of associated imaging data (raw and processed). The \acl{LaStBeRu}, dubbed \acs{LaStBeRu}, is the first data release of ``slcomp''.

The database comprises a compiled catalog with data on 31,569 unique objects from the literature, of which 6,660 galaxies and 4,612 clusters have been classified as lenses, and 10,421 galaxies and 831 quasars have been classified as source systems. Within the database, there are 3,822 systems with known source redshifts, 1,449 systems with measurements of Einstein radii, 29,791 lenses with reported magnitudes, and 1,385 sources with magnitudes in the bands mentioned above (as noted in Section~\ref{subsec:slfeatures}). In addition to the initial catalog, the database includes two additional cross-matching catalogs from photometric and spectroscopic surveys and a consolidated catalog that amalgamates information from prior catalogs and extends the available data. Table~\ref{tab:numbers} shows the increase in the cross-matching of identified data with other sources. By systematic cross-matching of spectrometric and photometric surveys, an increase of 27\% in redshifts and an increase of 851\% in velocity dispersion have been observed. Most of these data were obtained from \ac{SDSS}. The image catalog links all objects concatenated from previous catalogs, expanding the tabular data into a mosaic of raw and processed images. We obtained 366,905 squared image cutouts in \ac{FITS} format. This dataset includes 183,994 images of $20^{\prime\prime}$ size and 182,911 images of $4^{\prime}$ size, covering multiple photometric bands in both cases. Furthermore, we provide 91,920 \ac{RGB}-composed images processed with Trilogy and individual images for each photometric band. For completeness, we provide 125,215 images obtained from the \ac{LSB}, including images from \ac{SDSS}.

All four catalogs, including structured data from the literature, cross-matching photometry and spectroscopy, a consolidated catalog merging previous ones, and the catalog of raw and processed images, can be found in the ``slcomp'' repository, as stated in the Data Availability Section. We provide an exploratory dashboard displaying the complete data from the catalogs and processed cutouts, along with JuPyteR notebooks containing instructions on obtaining and filtering the data using queries based on references and original identifiers. The data is stored in an object storage service and accessible through the Python API.

\begin{table}
	\centering
	\caption{Cross-referencing the \acs{SL} compilation with photometric and spectroscopic catalogs. The second column displays the number of systems with the features from the first column available in the \acs{SL} compilation. The second and third columns are the number of matches identified in the photometric and spectroscopic catalogs, respectively. The final column exhibits the number of systems presenting novel data following the cross-referencing approach.}
	\label{tab:numbers}
    \resizebox{\columnwidth}{!}{
    \begin{tabular}{lrrrr}
        \hline
        & \ac{SL} compilation & Photo. catalogs & Spectro. catalogs & Novel \\
        \hline
        $z_L$ & 16520 & 16639 & 9006 & 4520 \\
        $\sigma_v$ & 595 & $\cdots$ & 5573 & 5070 \\
        $\mathrm{mag}_\mathrm{u}$ & 88 & 7533 & 7 & 7441 \\
        $\mathrm{mag}_\mathrm{g}$ & 8105 & 16995 & 7 & 9304 \\
        $\mathrm{mag}_\mathrm{r}$ & 8337 & 16900 & 28 & 9008 \\
        $\mathrm{mag}_\mathrm{i}$ & 4732 & 15661 & 2539 & 11366 \\
        $\mathrm{mag}_\mathrm{z}$ & 7600 & 16888 & 7 & 9525 \\
		\hline
	\end{tabular}}
\end{table}

\section{Applications} \label{sec:lastberu-applications}

\subsection{Modified Gravity Test using LaStBeRu Data} \label{sec:lastberu-applications-mod-grav}

\ac{SL} systems have been successfully used in many astrophysical and cosmological applications, such as constraining the cosmological parameters~\citep{Chen_et_al_2019} and the parameters of the dark energy equation of state~\citep{Cao_et_al._2015}, the total mass density profile~\citep{Bolton_et_al._2008, Chen_et_al_2019, 2019MNRAS.483.5649S, 2022MNRAS.517.3275E, 2023MNRAS.521.6005E, 2024MNRAS.52710480N}, and measurements of the $H_0$ parameter~\citep{2017MNRAS.468.2590S,2019MNRAS.490.1743C, 2020MNRAS.498.1420W}. One particularly unique test that can be carried out using \ac{SL} systems is to search for a specific signature of modified gravity, as shall be discussed in Section~\ref{sec:gammappn}. 

Most statistical applications of galaxy--galaxy \ac{SL} in cosmology require the knowledge of the following features: $z_L$, $z_S$, $\sigma_v$, and $\theta_E$ (and their uncertainties). The first three are typically obtained from spectroscopic data, while the last is derived from the \ac{SL} modeling of the systems.

In Figure~\ref{fig:feature-cuts}, we show the availability of these features from the \ac{LaStBeRu} compilation. The lines connecting the dots show the simultaneous availability of some of the features in the catalog, and the numbers at the top show the number of systems that have the selected features. We can see that 411 systems have all the required features from purely the archival data. For those or a sub-sample of them (see Section~\ref{sec:lastberu-cosmo-ground}), we can readily infer model parameters.

From this analysis, we can also identify systems with some missing features so that further work can be undertaken to complete the sample. For example, one may seek spectroscopic data to determine the source redshift, $z_S$, when it is missing from the catalog. Or one may obtain high \ac{SN} spectra to determine the lens velocity dispersion $\sigma_v$ or to improve available determinations, decreasing $\delta \sigma_v$. All these applications are discussed in Section~\ref{sec:lastberu-applications-follow}, where we describe spectroscopic follow-up projects to measure these quantities, selecting the targets from \ac{LaStBeRu}. Our group is also carrying out \ac{SL} modeling of the systems that do not have a determination of $\theta_E$ in the \ac{LaStBeRu} compilation, again to complete the availability of the features in the sample.

As mentioned before, in this work, we are particularly interested in what Rubin can achieve. Therefore, we focus on obtaining systems that are identifiable in ground-based data. The slcomp infrastructure allows one to efficiently perform this task by identifying systems with the required information from archival data and visually inspecting the selected ground-based data. This ensures the quality of the systems (e.g., good configurations for inverse modeling) and their suitability for the application (e.g., isolated lenses). To this end, in Section~\ref{sec:lastberu-cosmo-ground}, we compile a sample of systems with all relevant features that simultaneously show clear \ac{SL} features in the wide-field surveys from our compilation.

After creating a sample suitable for cosmological applications, we use this sample for estimating $\gamma_\mathrm{PPN}$, compare our results with the literature, and discuss their implications (Section~\ref{sec:gammappn}).

\subsubsection{Sample for Cosmological Applications} \label{sec:lastberu-cosmo-ground}

\begin{figure}
	\centering
	\includegraphics[width=\columnwidth]{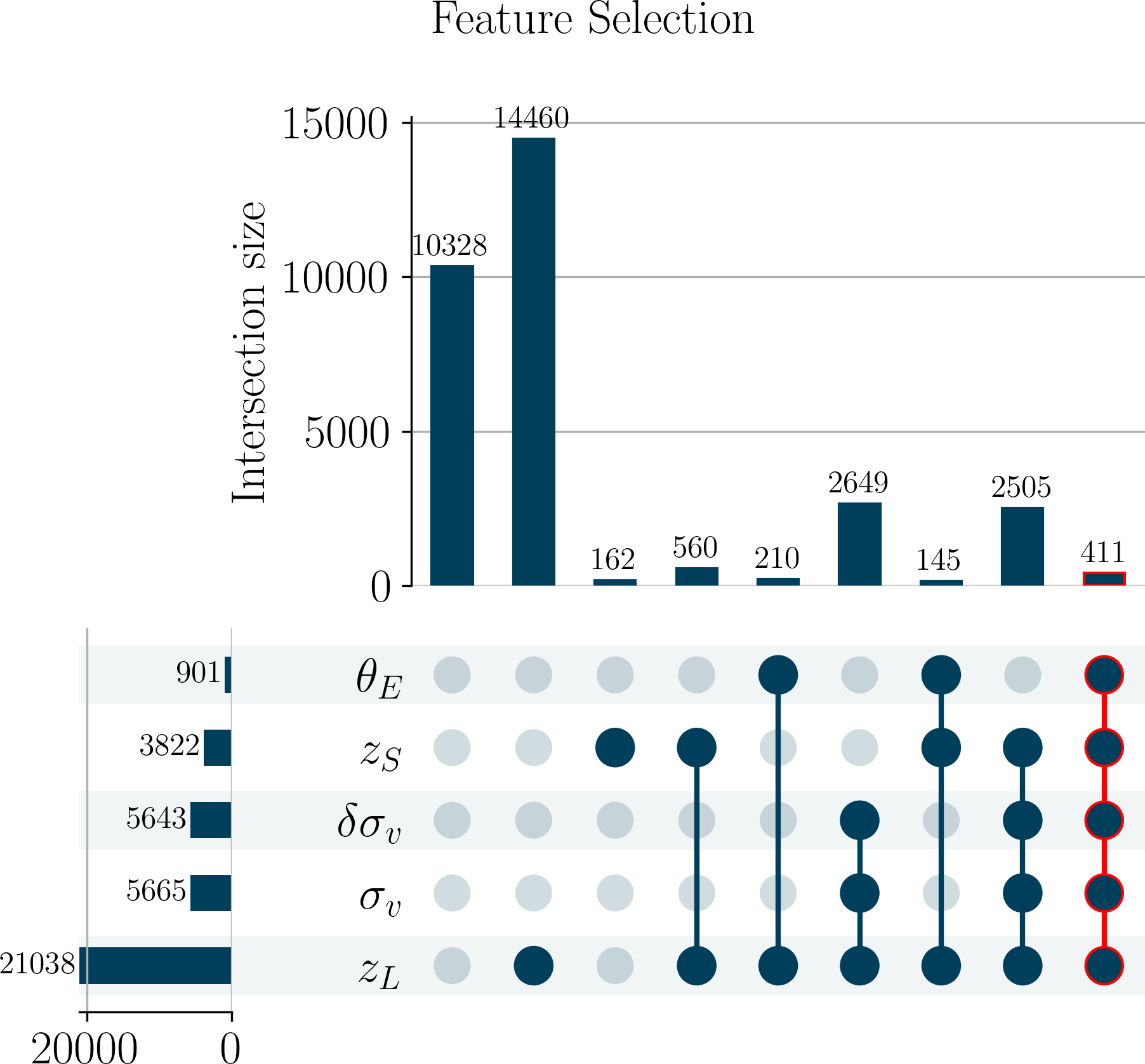}
	\caption{Availability of five parameters relevant for cosmological applications of \acs{SL} systems in the \acs{LaStBeRu} database. Each line in the lower panel indicates the systems for which the connected parameters are simultaneously available. Systems without intersections below 100 or empty values were omitted for simplicity.
 } \label{fig:feature-cuts}
\end{figure}

As mentioned above, several applications of \ac{SL} systems in statistical analyses require knowledge of the physical parameters of the system, especially $z_L$, $z_S$, $\sigma_v$, and $\theta_E$. The combination of these parameters is not readily available in the literature, as their determination does not share common conditions and methods. For example, the sources are usually very faint to allow for a determination of $z_S$ from spectra and typically require the presence of emission lines. So, only a fraction of the sources have their redshift determined. Measuring $\sigma_v$ requires a high \ac{SN} spectrum and moderate resolution, so not all lenses have this quantity determined from spectroscopy. Finally, only \ac{SL} systems with specific configurations (multiple peaked arcs, presence of counter images, etc.) enable a robust determination of $\theta_E$. Therefore, compiling a large number of systems is imperative to obtain a statistical sample with all needed physical parameters, which is enabled now with the \ac{LaStBeRu} dataset. Figure~\ref{fig:feature-cuts} shows the combined information available on \ac{LaStBeRu}, highlighting the presence of 411 systems with the parameters needed for the cosmological analyses.

To assemble a sample as homogeneous as possible, we selected only systems that had velocity dispersions determined from \ac{SDSS} spectroscopic data. This survey provides information from both the imaging and the spectroscopic data, including the seeing from the plates during spectroscopic observations, which is relevant for the modeling and has not typically been used in previous statistical analyses of this type. As we shall see, this information is relevant for the modeling and has not been used before in statistical analyses of the data of the type that will be discussed in this Section. A total of 352 systems (out of the 411) have $\sigma_v$ from \ac{SDSS}. However, not all data is valid. Several systems have precisely $\sigma_v=850$ km/s, indicating that this is a cap to identify outliers or bad fits. So, we remove systems with these values. Also, some systems do not have the seeing information (it is set to zero in the catalog) and are removed from the sample.

A further constraint on this sample is excluding systems classified as galaxy groups or clusters and those tagged as ``Non Single Lens Galaxy'' and ``Non Single Lens Galaxy Merged'' in the catalog. We also removed late-type galaxy lenses (e.g., \ac{SWEELS} catalog from~\citet{Treu_et_al._2011}). These restrictions are necessary to keep as much as possible only isolated \ac{ETG}, for which the modeling assumptions described in the next Section apply.

The visual inspection is also useful to eliminate further non-isolated lenses not tagged as such in the catalog compilation, such as the double galaxy systems SDSSCGB 8842.3 and SDSSCGB 8842.4. This emphasis on ground-based identifiable features is also key to mimicking the data quality and selection effects anticipated from \ac{LSST}.

To classify the systems, we used lensrater\footnote{\url{https://github.com/coljac/lensrater}}, in two tagging rounds to identify isolated \ac{ETG} in images with good quality: (i) using images from a single survey; (ii) a mosaic of available images from all surveys. Ultimately, only 206 systems passed all our criteria, leading to the sample we have dubbed as \textsc{LaStBeRu\_cosmo\_ground}.

\subsubsection{Testing Modified Gravity with Strong Lensing and Velocity Dispersions}\label{sec:gammappn}

In the cosmological context, a general metric perturbation in the \ac{FLRW} framework includes two scalar gravitational potentials, $\Phi$ (Newtonian) and $\Psi$ (curvature), rather than one. For linear scalar perturbations, the metric can be written as (see, e.g.,~\citet{2011RSPTA.369.4947B})
\begin{equation}
\mathrm{d}s^{2} = a(\tau)\left[-\left(1-2\Phi\right)d\tau^{2}+\left(1+2\Psi\right)\delta_{ij}dx^i dx^j\right]\,.
\label{eq:metric}
\end{equation}
The ratio of these potentials defines the slip parameter, $\eta \equiv \Psi/\Phi$, assumed here to be constant at relevant scales~\citep{2021PhRvD.104d4020T}. Nonrelativistic particle propagation is sensitive only to $\Phi$, while light propagation depends on both~\citep{2011RSPTA.369.4947B}. This distinction is key for testing gravity with \ac{SL} systems: $\Phi$ can be assessed through stellar dynamics in \ac{ETG} lenses, while \ac{SL} modeling probes the sum of the potentials. Under relevant assumptions, $\eta$ is equivalent to the \ac{PPN} parameter $\gamma_{\rm PPN}$~\citep{2021PhRvD.104d4020T}, the terminology used hereafter.

We focus on \ac{ETG} lenses, whose dynamics and profiles are relatively simple. Their mass density $\rho(r)$ and brightness $\nu(r)$ are modeled as power laws~\citep{binney1998}
\begin{equation} \label{Eq:mass_density_profile}
\rho(r) = \rho_0\left (\frac{r}{r_0}\right )^{-\alpha}
\end{equation}
and
\begin{equation} \label{Eq:brightness_profile}
\nu(r) = \nu_0\left (\frac{r}{r_0}\right )^{-\delta}.
\end{equation}
The dynamics of these collisionless systems (stars and Dark Matter) are described by the Jeans Equation, derived from the collisionless Boltzmann Equation. Assuming a constant anisotropy parameter $\beta = 1-\sigma_r^2/\sigma_t^2$ (where $\sigma_r$ and $\sigma_t$ are radial and tangential velocity dispersions), and using Equations~\eqref{Eq:mass_density_profile} and~\eqref{Eq:brightness_profile} in the spherical Jeans equation (derived from the metric in Equation~\ref{eq:metric}, neglecting cosmic expansion effects), the squared radial velocity dispersion is
\begin{equation} \label{Eq:Radial_Sigma_cte_beta}
\sigma^{2}_{r}(r)= \frac{GM(r)}{r} \frac{1}{\xi - 2\beta},
\end{equation}
where $\xi \equiv \alpha + \delta - 2$, and $M(r)$ is the dynamical mass.

Observationally, only the line-of-sight (\ac{LOS}) velocity dispersion, $\sigma^2_\mathrm{LOS}$, can be measured, typically as a brightness-weighted mean within a spectroscopic aperture. At a projected distance $r$ from the lens center, $\sigma^2_\mathrm{LOS}$ is given by~\citep{2006EAS....20..161K}
\begin{equation} \label{Eq:sigma_loss_cte_beta}
\sigma^2_\mathrm{LOS} = \frac{GM(r)}{r} \frac{\lambda(\xi) - \beta \lambda(\xi + 2)}{\lambda(\delta)(\xi - 2\beta_0)},
\end{equation}
where $\lambda(x) = \Gamma\left(\frac{x - 1}{2}\right)/\Gamma\left(\frac{x}{2}\right)$. To connect this to observations, we compute its mean within a circular spectroscopic aperture of radius $\theta_{\rm ap}$ (e.g., $1.5^{\prime\prime}$ for \ac{SDSS} I/II, $1^{\prime\prime}$ for BOSS). Atmospheric and telescope effects (seeing) blur the image, effectively convolving $\sigma^2_\mathrm{LOS}(r)$ with a Gaussian window function (variance $\sigma_\mathrm{atm}^2$)~\citep{Schwab_et_al_2010}, yielding
\begin{equation} \label{Eq:Avg_VD_cte_beta}
    \left \langle \sigma^2_\mathrm{LOS} \right \rangle =
    \frac{GM_E}{R_E} 
    \frac{\lambda(\xi)
   - \beta\lambda(\xi + 2)}{\lambda(\delta)(\xi - 2\beta)}\left(\frac{2\tilde{\sigma}^2_\mathrm{atm}}{\theta_\mathrm{E}^2}\right)^{(2-\alpha)/2} 
   \frac{\Gamma\left ( \frac{3 - \xi}{2} \right )}{\Gamma\left ( \frac{3 - \delta}{2} \right )},
\end{equation}
where $\tilde{\sigma}^2_\mathrm{atm} \approx \sigma^2_\mathrm{atm}(1 + \chi^2/4 + \chi^4/40)$ and $\chi = \theta_{\rm ap}/\sigma_\mathrm{atm}$. The square root of this quantity models the observed velocity dispersion.

To produce a more uniform observable, an aperture correction is applied to the measured velocity dispersion $\sigma_{\rm ap}$~\citep{Jorgensen}
\begin{equation}
    \sigma_{0} \equiv \sigma_{\rm ap}\left[\theta_{\mathrm{eff}}/\left(2\theta_{\rm ap}\right)\right]^{\zeta},
    \label{eq:Jorgensen}
\end{equation}
where $\theta_{\mathrm{eff}}$ is the effective (half-light) radius and $\zeta = -0.066\pm0.035$~\citep{Cappellari06}. For our analysis using the \textsc{LaStBeRu\_cosmo\_ground} sample, we use $\theta_{\mathrm{eff}}$ (Petrosian radius in $r$-band as baseline) and actual seeing $\sigma_\mathrm{atm}$ for each spectrum from \ac{SDSS} data (keywords \texttt{SEEING20}, \texttt{SEEING50}, \texttt{SEEING80} from `spPlate' table), a novel approach compared to previous works that assumed mean seeing values.

The connection to lensing analysis and modified gravity is made by relating this corrected observed velocity dispersion, $\sigma_0^{\mathrm{obs}}$ (derived from $\sigma_v$ using Equation~\eqref{eq:Jorgensen}), to a theoretical model $\sigma_{0}^{\mathrm{th}}$. This model incorporates the projected mass probed by lensing and the sensitivity of lensing to $\Phi+\Psi$. The model for $(\sigma_{0}^{\mathrm{th}})^2$ as a function of $\gamma_{\text{PPN}}$, mass/brightness distribution parameters, and the measured Einstein radius $\theta_E$ is~\citep{Schwab_et_al_2010, Cao_et_al_2017}
\begin{eqnarray}
    \label{eqn:sigma_model}
    \left(\sigma_{0}^{\mathrm{th}}\right)^2 & = & \left[\frac{2}{\left(1+\gamma_{\text{PPN}}\right)}\frac{c^{2}}{4}\frac{D_\mathrm{S}}{D_\mathrm{LS}}\theta_E\right]\frac{2}{\sqrt{\pi}}\left(\frac{2\tilde{\sigma}_{\mathrm{atm}}^{2}}{\theta_{E}}\right)^{2-\alpha} \\
    & & \times\left[\frac{\lambda\left(\xi\right)-\beta\lambda\left(\xi+2\right)}{\left(\xi-2\beta\right)\lambda\left(\alpha\right)\lambda\left(\delta\right)}\right]\frac{\Gamma\left(\frac{3-\xi}{2}\right)}{\Gamma\left(\frac{3-\delta}{2}\right)}. \nonumber
\end{eqnarray}
We aim to have $\sigma_{0}^{\mathrm{th}}(\theta_E,z_L,z_S,\theta_{ap},\sigma_\mathrm{atm},\mathbf{X}) \equiv \sigma_{0}^{\mathrm{obs}}$ (Equation~\eqref{eqn:sigma_model}), where observational inputs $\theta_E,z_L,z_S, \theta_\mathrm{eff}, \theta_{ap}, \sigma_\mathrm{atm}$ are known, and $\mathbf{X} = (\alpha, \delta,\beta,\gamma_\mathrm{PPN};\Omega_\Lambda, H_0)$ are the model parameters to be constrained. The cosmological parameter dependency enters through the distance ratio $D_S/D_{LS}$.

Model parameters are inferred using a likelihood function $\mathcal{L}\propto e^{-\chi^2/2}$ for the model in Equation~\eqref{eqn:sigma_model}~\citep{Liu_et_al_2022}. We assume global values for $\mathbf{X}$ across the sample. The uncertainty on $\sigma_{0}$, $\delta\sigma_{0}$, is modeled as~\citep{Liu_et_al_2022}
\begin{equation}
    \delta\sigma_{0}^{2} = \left\{ \delta\sigma_{\rm ap}^{2}/\sigma_{\rm ap}^{2}+\left[\ln\left(\theta_{\mathrm{eff}}/\left(2\theta_{\rm ap}\right)\right)\delta_\zeta\right]^{2}+0.03^{2}\right\} \sigma_{0}^{2},
\end{equation}
accounting for error propagation and an estimated 3\% systematic error from \ac{SL} modeling due to line-of-sight structures~\citep{2003AJ....125.1817B}.

Previous statistical measurements of $\gamma_{\mathrm{PPN}}$ using this method, such as from the \ac{SLACS} sample~\citep{Bolton_lenses, Schwab_et_al_2010} and larger compilations~\citep{Cao_et_al_2017, Liu_et_al_2022, Wei_et_al._2022}, yielded results consistent with \ac{GR} ($\gamma_{\mathrm{PPN}} \approx 1$). However, these often relied on \ac{HST} imaging or heterogeneous datasets, and many did not apply aperture corrections initially. Our motivation was to perform this test using the \textsc{LaStBeRu\_cosmo\_ground} sample (206 systems), focusing on systems with clear \ac{SL} features in ground-based data. Other studies using \ac{IFU} data for single systems~\citep{Collett, melo_carneiro_2023} or joint analyses with time-delays~\citep{yang_gamma_h0, liu_gamma_h0} also provide constraints on $\gamma_{\rm PPN}$, generally consistent despite different systematics.

For the \textsc{LaStBeRu\_cosmo\_ground} sample, we used the \texttt{emcee}~\citep{emcee} package for \ac{MCMC} analysis, assuming a flat $\Lambda$CDM cosmology (Planck20~\citep{2020A&A...641A...6P} parameters $\Omega_\Lambda, H_0$) and Gaussian priors: $\langle\alpha\rangle=2.00,~\sigma_\alpha=0.08$; $\langle\beta\rangle=0.18,~\sigma_\beta=0.13$; and $\langle\delta\rangle=2.40,~\sigma_\delta=0.11$. A key improvement was using system-specific seeing (from \ac{SDSS} spectra) and effective radii (Petrosian radii with \texttt{SEEING50} for baseline) in Equation~\eqref{eqn:sigma_model} and Equation~\eqref{eq:Jorgensen}. Our MCMC analysis (64 walkers, 15,000 steps, 500 burn-in) yielded $\gamma_{\mathrm{PPN}}=1.023\pm0.028$ from the \textsc{LaStBeRu\_cosmo\_ground} sample alone. Posteriors for $\beta$ and $\delta$ were prior-dominated. Results were robust to choices of effective radius type and seeing estimate, and consistent when neglecting aperture correction ($\gamma_{\mathrm{PPN}}=0.979\pm0.029$).

\begin{figure}
    \centering
    \includegraphics[width=\columnwidth]{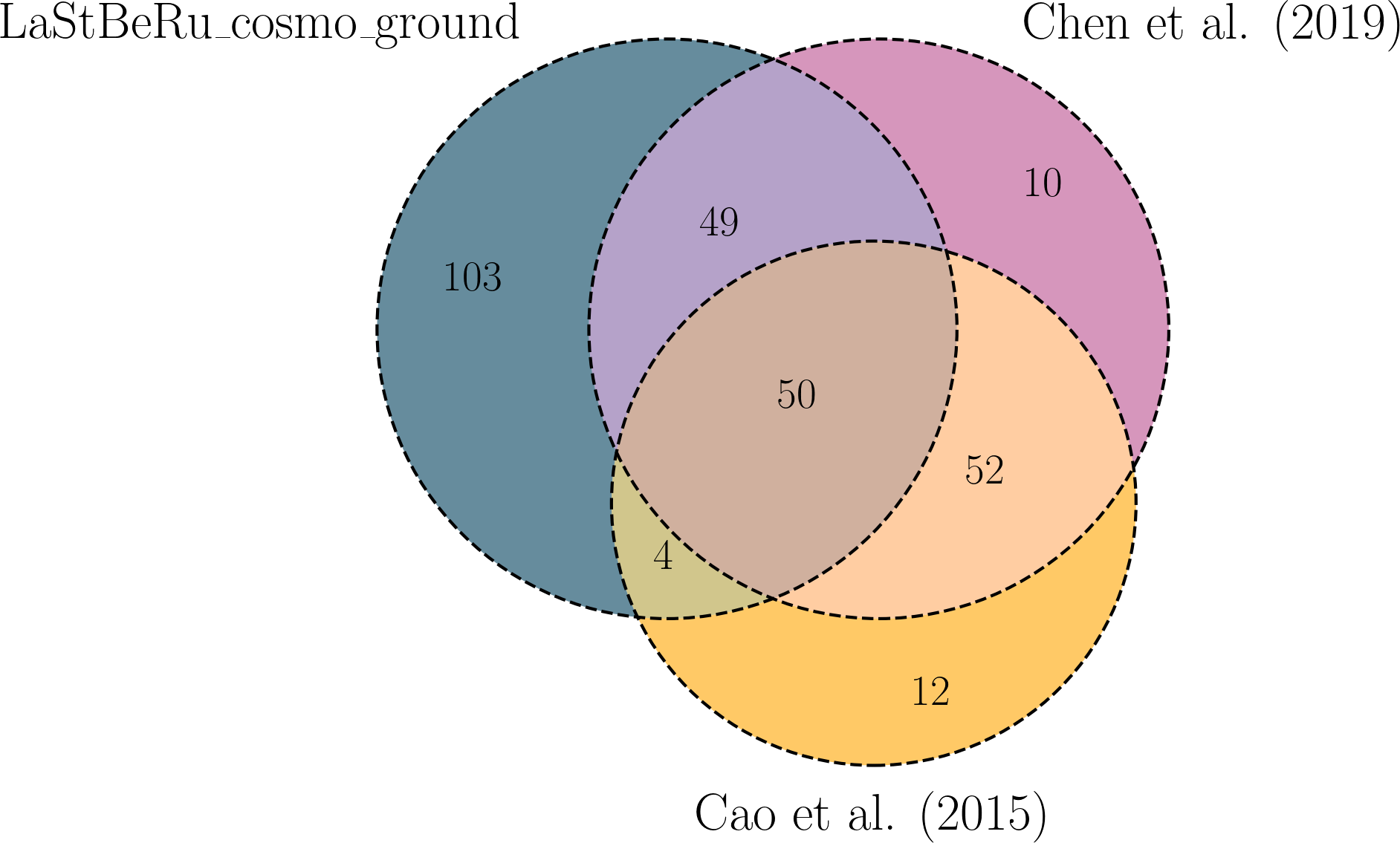}
    \caption{
 Intersections between the \textsc{LaStBeRu\_cosmo\_ground} sample and the samples compiled in~\citet{Cao_et_al._2015} and~\citet{Chen_et_al_2019}.} \label{fig:data-sample}
\end{figure}

\begin{figure}
    \centering
    \includegraphics[width=\columnwidth]{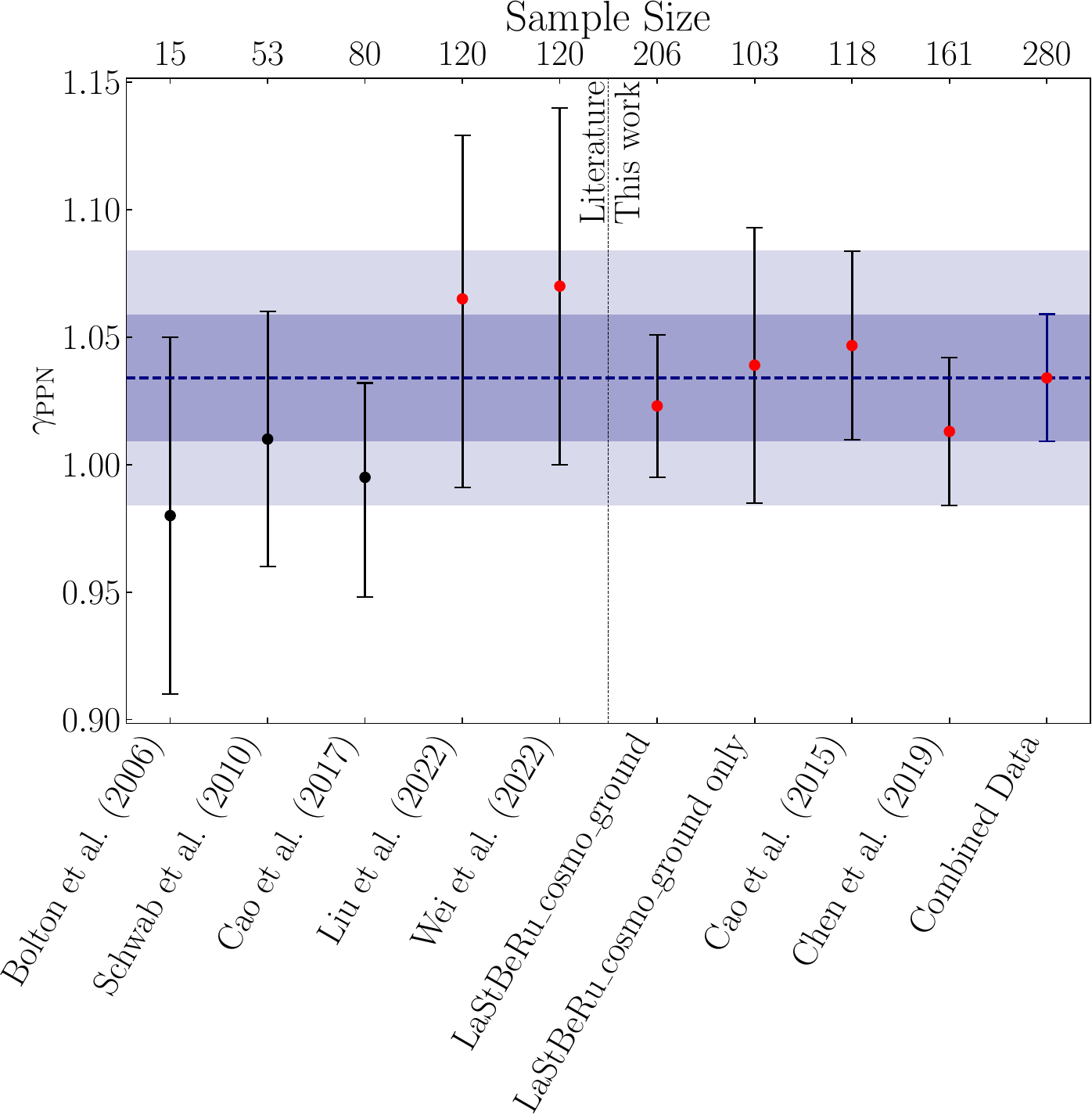}
    \caption{Estimated values for $\gamma_\mathrm{PPN}$ reported from the literature and from this work. The black dots represent results without aperture correction, while the red dots indicate results obtained, including this correction.} \label{fig:results}
\end{figure}

The \textsc{LaStBeRu\_cosmo\_ground} sample significantly differs from previous compilations by~\citet{Cao_et_al._2015} (derived from~\citet{Cao_et_al_2017}) and~\citet{Chen_et_al_2019} (used by~\citet{Liu_et_al_2022, Wei_et_al._2022}), primarily due to our visual inspection criteria for ground-based images and reliance on \ac{SDSS} data. As shown in Figure~\ref{fig:data-sample}, our sample includes 103 systems not previously used in such analyses. This independent subset yields $\gamma_{\mathrm{PPN}}=1.039\pm0.054$, consistent with our full sample and other literature results.
Combining all three catalogs (Figure~\ref{fig:data-sample}), totaling 280 unique \ac{SL} systems, without additional cuts on the literature samples, we find $\gamma_{\mathrm{PPN}}=1.034\pm0.025$. This is the tightest constraint on $\gamma_{\mathrm{PPN}}$ from this type of analysis to date and is consistent with \ac{GR} within two standard deviations. Figure~\ref{fig:results} summarizes these results and compares them with literature values.

In conclusion, using the \ac{LaStBeRu} compilation, we constructed the \textsc{LaStBeRu\_cosmo\_ground} sample of 206 \ac{ETG} lens systems suitable for cosmological analysis, mimicking next-generation ground-based survey data like \ac{LSST}. This sample, unique in its reliance on ground-based data and homogeneous \ac{SDSS} spectroscopy (with object-specific seeing and radii), provides competitive and independent constraints on $\gamma_\mathrm{PPN}$. Our ``almost blind'' analysis approach, where selections were predefined based on data quality, enhances confidence in the results. This not only yields some of the most stringent constraints on $\gamma_\mathrm{PPN}$ but also demonstrates the utility of the \ac{LaStBeRu} compilation for other cosmological studies. The analysis pipeline is publicly available in the \href{https://github.com/CosmoObs/SLtools/blob/master/sltools/rings2cosmo/rings2cosmo.py}{SLtools repository}.

\subsection{Follow-Up Observations} \label{sec:lastberu-applications-follow}

As discussed in Section~\ref{sec:lastberu}, the \ac{LaStBeRu} compilation contains several physical features of the \ac{SL} systems that are relevant for astrophysical and cosmological applications. One may use this purely archival data for applications, as carried out in Section~\ref{sec:gammappn} for the $\gamma_{\mathrm{PPN}}$ parameter. One can also identify missing quantities, such as source redshifts crucial for precise lens modeling or improved velocity dispersions, to reduce uncertainties in $\gamma_{\mathrm{PPN}}$ estimates targeting these specific systems, i.e., with follow-up observations. For example, we can seek spectroscopic determinations of the lens and source redshifts, obtain higher \ac{SN} spectra for measuring stellar velocity dispersions, or acquire better-quality images to perform the inverse modeling.

Rather than completing available information in the sample, we can seek new observational data outside the scope of \ac{LaStBeRu}. For example, for several applications, it is crucial to obtain spatially resolved spectroscopy across the whole image plane (i.e., containing images, lens components, and structure along the \ac{LOS}), which is carried out using \ac{IFU}. The slcomp framework assists in selecting appropriate systems for such follow-up, taking into account criteria like redshift ranges, ensuring that emission lines are present in the sources, and select systems with the desired lens configuration that fit the field-of-view of the instrument. It also helps define observational constraints such as \ac{RA} and \ac{Dec} range for suitable observing conditions, and minimum brightness to achieve desired \ac{SN}, often narrowing down the candidate pool significantly.

Tabular data alone is generally insufficient to appropriately select \ac{SL} targets. One needs actual images to assess the quality of the systems and their suitability for the problem at hand. For example, we may require specific features for the lens inversion, such as the presence of multiple peaks in arcs and counter-images, or the lens and its environment, such as being an isolated \ac{ETG}, avoiding the presence of foreground or background galaxies, which could contaminate the light of the images and influence the \ac{SL} modeling. The images may also be required to design the observations, such as determining the orientation of the spectroscope slit or checking if the systems fit in a given \ac{IFU}.

As discussed in Section~\ref{sec:lastberu-applications-mod-grav}, we are particularly interested in performing modified gravity tests from the combination of \ac{SL} data and the velocity dispersions of lens galaxies. The uncertainty in $\sigma_v$ is the dominant source of statistical uncertainty in this kind of analysis. As the systems from \textsc{LaStBeRu\_cosmo\_ground} have \ac{SDSS} spectroscopy, we sought to obtain higher \ac{SN} spectra to improve the determination of $\sigma_v$ for a few selected systems. More generally, we wanted to test the ability to obtain $\sigma_v$ from targeted follow-up observations and try to determine $z_S$ for some of the sources. To this end, we prepared observational proposals for the \ac{SOAR}, Gemini, and Jorge Sahade/\ac{CASLEO} telescope using the slcomp framework. Using the $\sigma_v$ measurements of the objects observed with the \ac{SOAR} and Gemini telescopes, we obtained the first constraint on $\gamma_{\mathrm{PPN}}$ from ground-based data through a self-consistent, end-to-end analysis that will be presented in França et al., in preparation. The purpose of this Section is to briefly show how the slcomp framework supports the selection and planning of \ac{SL} targets for follow-up campaigns, as detailed in Sections~\ref{sec:observational-plan} and~\ref{sec:proposals}.

\subsubsection{Observational Plan} \label{sec:observational-plan}

To design an observational plan to follow up some known \ac{SL} systems, we start with the \ac{LaStBeRu} compilation and filter the targets by the range of \ac{RA} and \ac{Dec} suitable for the chosen telescope in the particular observation window. Next, we determine the maximum magnitude allowed by our targets, taking into account the desired minimum \ac{SN} we want to achieve on a specific measurement with reasonable exposure times (a few hours at most, given the observability of the object along the night). This already shows the power of using a consolidated catalog, as the original \ac{SL} catalog may not contain magnitudes for the targets (lens or source) or may not be in the desired filter. With the \ac{LaStBeRu}, we have cross-matched magnitudes in several bands for most lenses and a few images. This allows us to make a more uniform cut in magnitude, even if we use systems from different \ac{SL} surveys. We often want to select the targets in a given lens and/or source redshift range. Again, thanks to the cross-match, one has more values of spectroscopic redshifts for the systems than in the original \ac{SL} catalogs and photometric redshifts for most lenses and some of the images.

Determining source redshifts is one of the strongest limiting factors in \ac{SL} observations since the images are usually very faint. Without emission lines, it becomes almost impossible to obtain $z_S$. Therefore, the yield to determine $z_S$ in spectroscopic follow-up programs is usually low. Consequently, if we want to study specific system properties, such as a high \ac{SN} measurement of the lens velocity dispersion, we must ensure targeting systems with known $z_S$. Again, selecting systems with known $z_S$ (and $z_L$) becomes trivial when using slcomp.

After filtering from the tabular data with constraints from a particular science goal, retrieving all available image cutouts in \ac{FITS} format or as \ac{RGB} color composites was easy. The latter is useful for visual inspection. At the same time, the former is essential to design the observations (e.g., slit or \ac{IFU} position) and to carry out the scientific analysis of the image (for example, for inverse modeling). If the cutout is initially unavailable from the surveys we used in our database, we already included images from \ac{LSB}.

\begin{figure*}
	\centering
	\includegraphics[width = \textwidth]{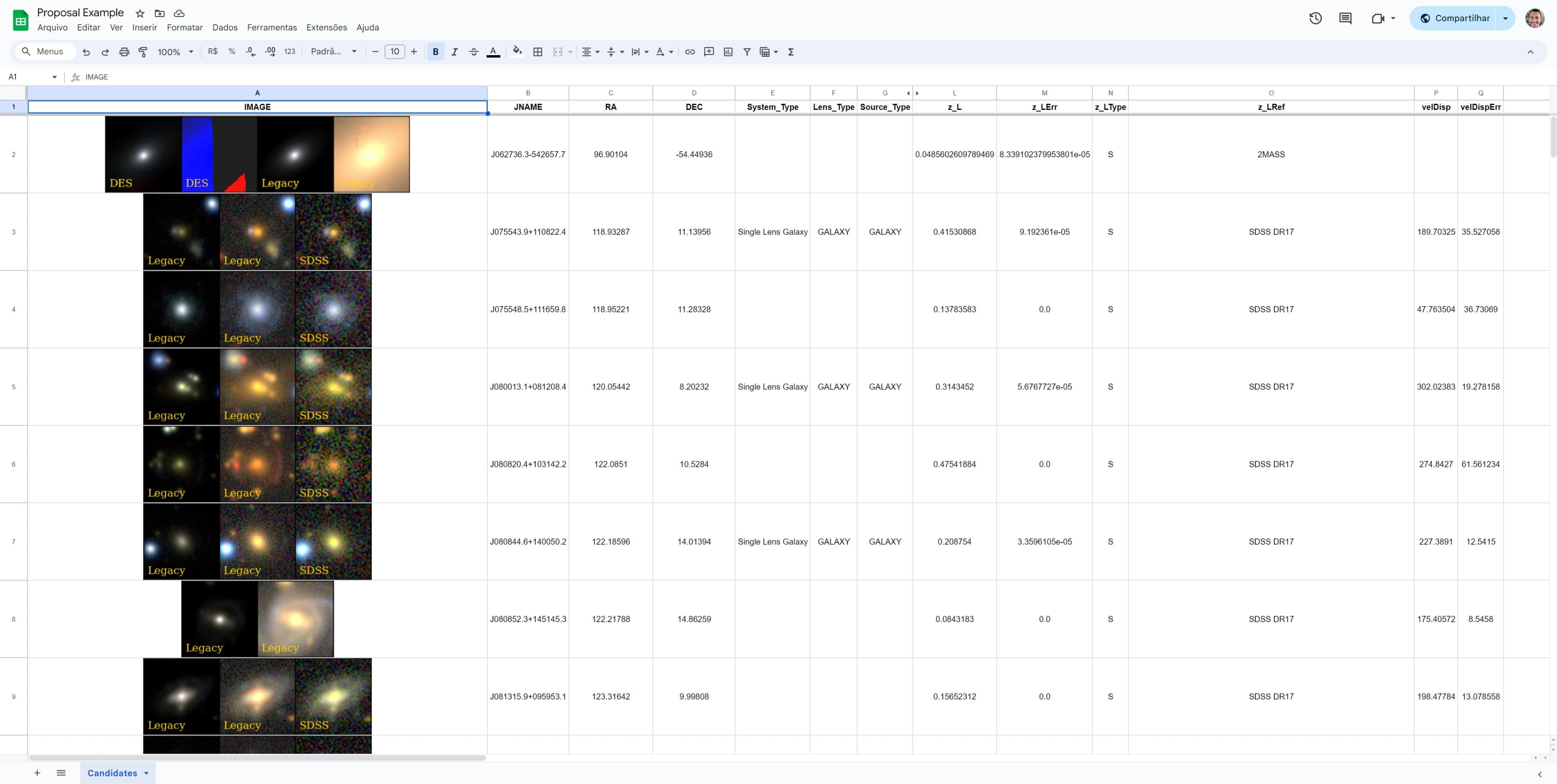}
	\caption{Example of a shared spreadsheet with data from \acs{LaStBeRu} and each available mosaic of cutouts for the observational candidates.}
	\label{fig:spreadsheet}
\end{figure*}

The products of these steps of target selection are a spreadsheet containing all tabular data available in the \ac{LaStBeRu} database and a collection of cutouts for later visual inspection. For the reader interested in this process, we provide a \href{https://github.com/CosmoObs/slcomp/blob/1c057c8f11badadee617410cb05058e0d3a64819/notebooks/proposals/proposals.ipynb}{JuPyteR notebook} to illustrate the target selection process for follow-up observations. To facilitate collaboration, we shared data with team members using a Google Spreadsheet (see Figure~\ref{fig:spreadsheet}) and uploaded all image mosaics to GitHub.

This greatly simplifies the task of manually selecting and ranking the systems with the selected information in the spreadsheet, allowing the insertion of comments in the same document. This can be useful during the target selection process and observations to insert information for each system, such as date, time, and exposure time.

\subsubsection{Observational Proposals} \label{sec:proposals}

We conducted an observational proposal to \ac{CASLEO} (CASLEO 2022A) aimed to target bright \ac{SL} systems from the \ac{LaStBeRu} sample without available source redshift measurements, selecting candidates likely to exhibit detectable emission lines within the spectral range of the \ac{REOSC} spectrograph at the Jorge Sahade 2.15 m telescope (\ac{CASLEO}). A control sample of known lenses with emission lines from the \ac{SILO} catalogue was also included to validate the observational setup. Out of 1333 accessible systems, we selected 26 \ac{SL} candidate systems based on brightness, redshift range, and the presence of clear lensing features. Observations of ten systems were conducted in March 2022, however, all resulting spectra presented low \ac{SN}, showing no promising lines from which to obtain lens and source redshifts. This outcome is likely due to a combination of bad seeing, the small instrument aperture, and low \ac{CCD} throughout. Since then, the primary mirror of the telescope was aluminized, and the \ac{REOSC} \ac{CCD} was changed by a more efficient one. Therefore, we plan to repeat the observations eventually. We also prepared an \ac{IFU} follow-up proposal to the Gemini telescope to constrain $\gamma_\mathrm{PPN}$ using 3 selected \ac{SL} systems, chosen based on redshift availability, brightness, lens configuration, and compatibility with \ac{GMOS}-\ac{IFU} observations. Unfortunately, this \ac{IFU} proposal did not reach enough priority to be observed. In the following paragraphs, we briefly describe the follow-up proposals submitted to the Gemini (program GN-2021B-FT-214) and SOAR (programs SO2022A-020 and SO2022B-018) telescopes, which employed the slcomp infrastructure for target selection. These efforts resulted in a subsample of \ac{SL} candidate systems for modified gravity tests, with measurements of $\sigma_v$, $z_L$, and $z_S$. A more detailed description of the selection of the objects and the corresponding $\gamma_{\mathrm{PPN}}$ constraint with the \ac{SOAR} and Gemini follow-up data will be presented in França et al., in preparation.

We submitted an observation proposal as a feasibility study using the Gemini Telescope, under the Fast Turnaround queue, entitled ``Feasibility study for testing GR with a new strong lensing system'' (program GN-2021B-FT-214). This proposal aimed to obtain the lens velocity dispersion ($\sigma_v$) and as a byproduct confirm the \ac{SL} candidates by measuring the image redshifts. The motivation was to assess the capability of making accurate and precise determinations of $\sigma_v$ to be used in gravity tests that combine the dynamics of the lens and the \ac{SL} modeling to set constraints on the so-called $\gamma_{\mathrm{PPN}}$ parameter. In Section~\ref{sec:lastberu-applications-mod-grav}, we provided more information on how this parameter is determined from the data.

Using data and image cutouts from the \ac{LaStBeRu} database, we filtered the systems based on their observability during this call between December (2021) and February (2022). Additionally, we required the systems to have known lens and source redshifts, either photometric or spectroscopic. This restriction led to a sample of 1069 candidates.Using the Google spreadsheet feature (Figure~\ref{fig:spreadsheet}), we visually inspected these systems in the images of all surveys available in the \ac{LaStBeRu} catalog for cutouts. The chosen systems must show a clear \ac{SL} morphology compatible with images from a single elliptical lens to enable accurate inverse reconstruction of the lens mass distribution using lensing techniques. Given the available time for this Fast Turnaround proposal, we observed only one system, namely J083933-014044. The spectroscopic analysis of this system and the corresponding modeling results will be reported and discussed in França et al., in preparation.

In the SOAR follow-up programs (SO2022A-020 and SO2022B-018), we submitted two proposals titled ``Testing general relativity in galaxies with combined velocity dispersion and gravitational lensing''. The main objective was to measure the velocity dispersions of lens galaxies from the width of emission lines in the spectra. As a byproduct, some spectroscopic redshifts for the lens and the source might have been obtained, allowing us to compare them with the available data from LaStBeRu compilation. We were awarded 6 nights in 2022A (SO2022A-020) and 5 nights in 2022B (SO2022B-018) to observe \ac{SL} systems with the Goodman spectrograph at the \ac{SOAR} telescope. The purpose of these proposals are threefold: i) to obtain a new measurement of $\gamma_{\mathrm{PPN}}$ from an independent dataset, ii) to evaluate the improvement in the determinations of velocity dispersion with the proposed \ac{SOAR} setup, and iii) to assess the use of \ac{SOAR} as a follow-up instrument to test modified gravity in the context of the next-generation imaging surveys.

\begin{figure}
    \centering
    \includegraphics[width=\columnwidth]{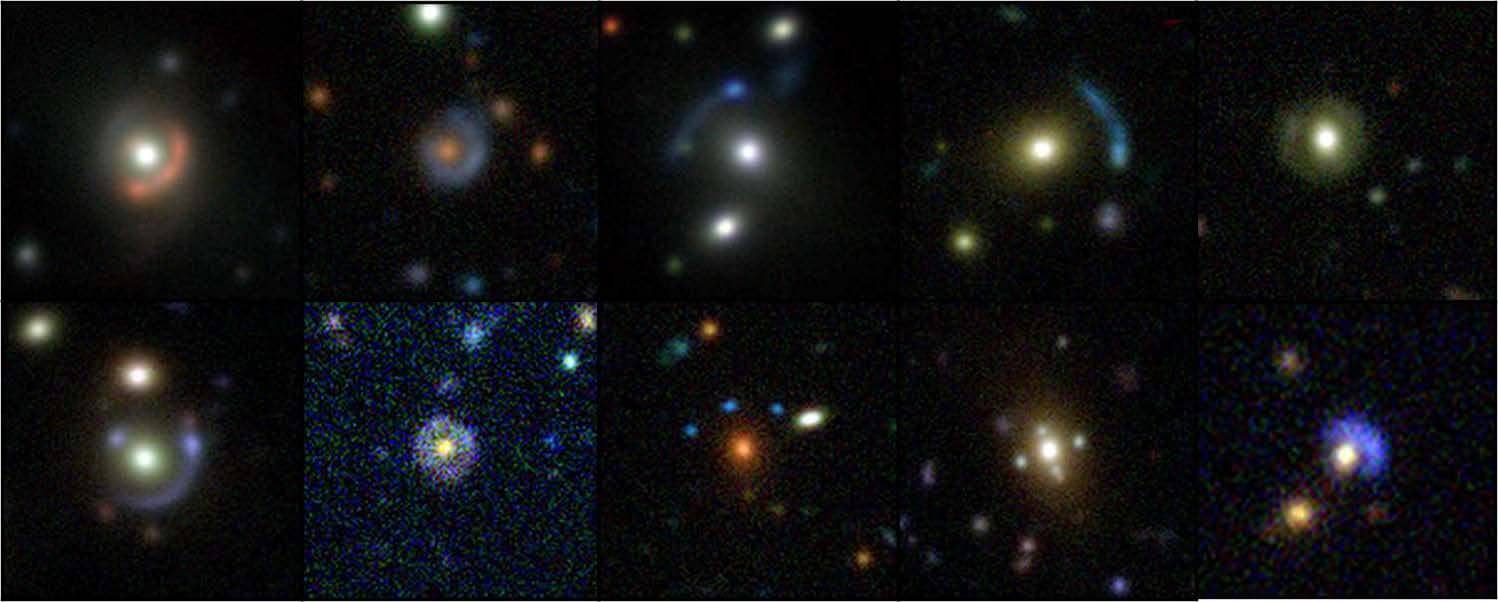} 
    \caption{10 \ac{SL} candidate systems from the selected subsample. All exhibit characteristics that may resemble SL features in the \ac{HSC} data, such as multiple images and multiply peaked gravitational arcs or rings, which allow for the inverse modeling of these systems to determine the Einstein radius reliably.}
    \label{fig:2022A-fig-1}
\end{figure}

We ultimately identified 30 galaxies observable from the Southern Hemisphere, near the celestial equator, deemed suitable for the first proposal (SO2022A-020). These systems were selected from a broader sample of \ac{SL} candidates based on having known spectroscopic redshifts for both the lens ($z_L$) and the source ($z_S$), as well as through visual inspection to retain only those with high-quality imaging and morphologies that allow for a precise \ac{SL} reconstruction of their mass distribution. We have chosen objects spanning the \ac{RA} range from 7h to 18h to achieve greater flexibility for scheduling throughout the semester. We restricted the systems in the range $-70\leq\mathrm{Dec}\leq+10$ to ensure observability, prioritizing targets with $\mathrm{Dec}\leq-30$ due to the lack of windscreen at \ac{SOAR}. Also, targets are more concentrated on low \ac{RA} due to \ac{SDSS} coverage. Out of the 30 initially selected systems, 12 were observed. Some examples of selected systems are shown in Figure~\ref{fig:2022A-fig-1}.

The results from the 2022A campaign were promising, but we could not observe the total number of systems we intended. Therefore, we submitted a proposal for 2022B to complete our sample. Using a similar selection criteria, we chose 17 galaxies observable from the Southern hemisphere with good observability for semester 2022B. These systems complement the 12 observed in the 2022A sample by spanning a larger redshift interval, particularly extending the sample to higher redshifts. Out of the 17 selected systems selected for the 2022B, 11 were observed. The analysis of the spectroscopic data, measurements of the lens stellar velocity dispersion $\sigma_v$ for the SOAR observed systems, and the analysis of the corresponding imaging data with inverse \ac{SL} modeling will be also reported and discussed in França et al., in preparation. 

These follow-up proposals highlight how the slcomp framework and LaStBeRu database enable targeted, efficient follow-up observations that can be used for \ac{SL} science.

\section{Summary and Concluding Remarks} \label{sec:conclusions}

Upcoming wide-field surveys, such as Rubin \ac{LSST}, Euclid, and Roman, will increase the number of known strong lenses by one or two orders of magnitude, leading to hundreds of thousands of systems. Predicting their abundance, deriving their astrophysical and cosmological applications, and modeling individual systems will require fast numerical approaches. As stated above, this is one of the motivations to search for analytical expressions, for example, to enable an efficient probing of parameter spaces and generate many simulated systems. Another approach is to train \ac{ML} methods, which can perform several fast prediction and modeling (regression) tasks. These usually require massive datasets to train and validate the methods. Moreover, these models often require training data that closely resembles the actual data where they will be applied. Regardless of the chosen approach, it is crucial to be able to test all kinds of analyses on real data. 

To this end, and motivated by the myriad applications of \ac{SL}, we have developed a framework dubbed ``slcomp'' to compile, cross-match, and aggregate information on \ac{SL} systems, generating cutouts of these systems on the current wide-field surveys. The development of this framework, together with its derived data products, is described in Section~\ref{sec:lastberu}.

With minimal human intervention, the database can be updated with new searches for \ac{SL} systems. The database contains a compilation of previously reported lensing systems, ranging from serendipitous to automated searches in wide-field surveys. We compile the available information in the source \ac{SL} catalogs and identify duplicates (i.e., the same system across different surveys or papers), keeping all available information for each system. We also cross-match the resulting catalog with photometric and spectroscopic surveys. The resulting product is a consolidated database gathering data from the \ac{SL} catalogs, and the cross matches with spectroscopic and photometric catalogs (including redshifts when available). We keep the values for each physical parameter (e.g., redshift, Einstein radius) from each catalog and their provenance. To simplify the practical applications of the catalog, we also assign single values to each physical quantity, following the quality criteria described in Section~\ref{sec:consolidated}.

In addition to the consolidated database, we produce image cutouts for each system in several wide-field imaging surveys. The cutouts are performed in all bands available in each survey, and they were made in two sizes: $20^{\prime\prime}$ and $4^{\prime}$ on a side corresponding to scales suitable for galaxy and cluster \ac{SL}, respectively. Composite (\ac{RGB}) color images are also produced for each system in each survey. This dataset has many applications, but one of the main drivers is to help prepare for the upcoming Rubin \ac{LSST} data. Furthermore, all our cutouts are from ground-based data, and most surveys have a \ac{PSF} that will be comparable to the \ac{LSST} one (subarcsecond seeing). To our knowledge, this represents the most extensive compilation of tabular data and ground-based imaging for known lenses achievable before the start of Rubin \ac{LSST} operations. We hope this dataset can be fully exploited before and as a preparation for the Rubin data. For this reason, we have dubbed this current database version as \acl{LaStBeRu} (\acs{LaStBeRu}).

Historically, this project started by seeking suitable targets for observational programs for spectroscopic follow-up of \ac{SL} systems. In particular, we were looking for systems with known $z_L$ and $z_S$ but with poor or no measurements of the stellar velocity dispersion, which were bright enough to reach the necessary \ac{SN} for spectroscopy on a given instrument/telescope and that could be adequately modeled. This required isolated galaxy--galaxy lenses (not groups of clusters, no close neighbors) with multiple images (or multiple brightness peaks). Furthermore, the systems needed to be visible in a given region of the celestial sphere with good observing conditions for a given semester in a given telescope location. These conditions are challenging to meet given the number of targets available in a single \ac{SL} catalog. Therefore, to carry out the selection necessary for the observational programs, we needed to collect systems from many catalogs, cross-match with known redshift and photometric catalogs, and be able to visually inspect the systems to ensure their suitability to be modeled on the ground-based data, both in terms of the lens (isolated \ac{ETG}) and images (typical morphology of lensing by an elliptical galaxy). The need to meet all these conditions and to be useful for selecting targets for other proposals has led to the slcomp framework. The database is accessible through the Python API, and we provide JuPyteR notebooks with examples of how to interact with the database, as mentioned in Section~\ref{sec:lastberu}.

As discussed in Section~\ref{sec:lastberu-applications-mod-grav}, one of the motivations for our \ac{SL} work was to perform a test of modified gravity in terms of the \ac{PPN} parameter. This can be done by combining the \ac{SL} modeling with the dynamics of the lens, which can be coarsely determined by its \ac{LOS} stellar velocity dispersion, as reviewed in Section~\ref{sec:gammappn}. This analysis can be readily carried out using the archival data compiled in the \ac{LaStBeRu} sample. We used the slcomp framework to select the systems with the needed physical parameters from the catalog. We kept only those with the lens velocity dispersion from \ac{SDSS} data for uniformity. We further required that all systems be clearly identifiable in ground-based images. This yielded a new sample, the \textsc{LaStBeRu\_cosmo\_ground}, which is suitable for other cosmological applications. From this sample, we were able to: i) obtain for the first time constraints on $\gamma_\mathrm{PPN}$ that is totally independent of previous samples, ii) use only data with \ac{SL} features identified in ground-based images, and iii) combine it with previously available samples to obtain the most stringent constraints achieved so far on $\gamma_\mathrm{PPN}$ through this type of analysis. We believe our results are less affected by systematic effects than the previous studies due to the uniformity of the data from the lens, including its effective radius and the spectroscopic data.

We used the \ac{LaStBeRu} compilation and slcomp framework to select systems for spectroscopic follow-up aiming to determine the velocity dispersion of the lens, as discussed in Section~\ref{sec:lastberu-applications-follow}. This led to successful proposals using data to provide the first end-to-end determination of $\gamma_\mathrm{PPN}$. For the first time, we used priors on the brightness and density slopes from the same data, providing a more self-consistent analysis compared to previous works. With this analysis, we will derive the first determination of $\gamma_{\rm PPN}$ fully from ground-based data. The results will be presented in França et al., in prep.

As mentioned above, the slcomp database can be continuously updated, and we plan to incorporate data from \ac{DELVE}, \ac{Pan-STARRS}, and \ac{HSCLA}. We will include Rubin \ac{LSST} data as soon as the first images and \ac{SL} samples become publicly available. We also plan to extend the compilation to include images from space-based observatories, such as \ac{HST}, \ac{JWST}, Euclid, and Roman. Another possibility to increase the data is merging or incorporating \ac{LaStBeRu} with public databases, such as lenscat, and other frameworks.

In essence, the framework and database presented here serve as a ready-to-use platform for many applications. Researchers needing a watchlist for transient events in lensed systems, a clean sample for cosmological analyses, or specific targets for observational follow-up can now query our database, retrieve pre-processed cutouts, and make their selections efficiently. For the machine learning community, the forthcoming visually inspected and tagged version of \ac{LaStBeRu} will provide a new benchmark dataset for training and testing classification and regression algorithms, as will the large, uniformly modeled samples that will be produced.

\section*{Acknowledgements}

This work was partially funded by the Coordenação de Aperfeiçoamento de Pessoal de Nível Superior – Brazil (CAPES). JPCF acknowledges support from the Conselho Nacional de Desenvolvimento Científico e Tecnológico - Brazil (CNPq - 140210/2021-0). MM acknowledges support from AGENCIA I+D+i - Argentina (project PICT-2021-GRF-TI-00816), CNPq (316239/2023-2), and the State of Rio de Janeiro (FAPERJ - E-26/202.687/2019 and E-26/210.079/2020). The authors thank the developers of Python~\citep{10.5555/1593511} (including standard libraries--argparse, collections, gc, getpass, glob, io, itertools, json, multiprocessing, os, pathlib, re, requests, shutil, subprocess, sys, urllib, warnings), Astropy~\citep{astropy}, Cython~\citep{behnel2011cython}, emcee~\citep{emcee}, GetDist~\citep{2019arXiv191013970L}, healpy~\citep{2020ascl.soft08022Z}, Joblib~\citep{joblib}, matplotlib\_venn~\citep{matplotlibvenn}, Matplotlib~\citep{Hunter:2007}, MOCPy~\citep{2019ASPC..521..487B}, NumPy~\citep{2020NumPy-Array}, pandarallel~\citep{pandarallel}, pandas~\citep{mckinney2010data}, parmap~\citep{parmap}, Pillow~\citep{andrew_murray_2023_8349181}, pqdm~\citep{pqdm}, pyvenn~\citep{pyvenn}, PyVO~\citep{2014ascl.soft02004G}, SciPy~\citep{2020SciPy-NMeth}, SEP~\citep{Barbary2016, 1996A&AS..117..393B}, tqdm~\citep{casper_da_costa_luis_2023_8233425}, Trilogy~\citep{Trilogy}, UpSetPlot~\citep{6876017}, xmatch~\citep{xmatch}. This work made use of the CHE cluster, managed and funded by COSMO/CBPF/MCTI, with financial support from CNPq, FINEP and FAPERJ. The authors would like to acknowledge the use of the computational resources provided by the Sci-Com Lab of the Department of Physics at UFES, which was funded by FAPES and CNPq. This research uses services or data provided by the Astro Data Lab at NSF's NOIRLab. NOIRLab is operated by the Association of Universities for Research in Astronomy (AURA), Inc. under a cooperative agreement with the National Science Foundation. This research has made use of the VizieR catalogue access tool, CDS, Strasbourg, France. This research is based on data collected at the Subaru Telescope and retrieved from the HSC data archive system, which is operated by the Subaru Telescope and Astronomy Data Center (ADC) at NAOJ. Data analysis was in part carried out with the cooperation of Center for Computational Astrophysics (CfCA), NAOJ. We are honored and grateful for the opportunity of observing the Universe from Maunakea, which has the cultural, historical and natural significance in Hawaii. This research includes data that has been provided by AAO Data Central.


\section*{Data Availability}

The ``slcomp'' framework and the \ac{LaStBeRu} database version, which includes all the data access documentation, are available at \url{https://github.com/CosmoObs/slcomp}.



\bibliographystyle{mnras}
\bibliography{reference} 


\bsp	
\label{lastpage}
\end{document}